\documentclass[11pt,a4paper]{article}
\usepackage{jheppub}
\usepackage{amsmath,amstext,amssymb}
\usepackage{graphicx}
\usepackage{feynmp}

\def\N4{${\cal N} = 4$}
\def\nn{\nonumber}
\def\Yf{{\rm Y}}

\title{Multi-Regge Limit of the n-Gluon Bubble Ansatz}
\subheader{DESY 12-122}
\author[a]{J.\ Bartels,}
\author[b]{V.\ Schomerus}
\author[b]{and M.\ Sprenger}

\affiliation[a]{II.\ Institute for Theoretical Physics,\\Hamburg University, Germany}
\affiliation[b]{DESY Theory Group,\\Hamburg, Germany}

\emailAdd{joachim.bartels@desy.de}
\emailAdd{volker.schomerus@desy.de}
\emailAdd{martin.sprenger@desy.de}

\abstract{We investigate n-gluon scattering amplitudes in the multi-Regge
region of \mbox{\N4} supersymmetric Yang-Mills theory at strong coupling. Through
a careful analysis of the thermodynamic bubble ansatz (TBA) for surfaces in
$AdS_5$ with n-g(lu)on boundary conditions we demonstrate that the multi-Regge
limit probes the large volume regime of the TBA. In reaching the multi-Regge
regime we encounter wall-crossing in the TBA for all $n>6$. Our results
imply that there exists an auxiliary system of algebraic Bethe ansatz equations
which encode valuable information on the analytical structure of amplitudes at
strong coupling.}


\begin{document}

\newcommand{\Y}[3]{\mathrm{Y}_{#1,#2}^{\left[#3\right]}}
\newcommand{\Ynt}[3]{\mathit{Y}_{#1,#2}^{\left[#3\right]}}
\newcommand{\U}[2]{\frac{\Ynt{2}{#1}{#2}}{1+\Ynt{2}{#1}{#2}}}

\maketitle

\section{Introduction}

The computation of gluon scattering amplitudes in gauge theories such as
Quantum Chromodynamics (QCD) or its supersymmetric cousins is a daunting task.
Over the last few years much progress has been made in the context of \N4
supersymmetric Yang-Mills (SYM) theory, both at weak and strong coupling.
These exciting developments exploit new hidden symmetries, such as
dual conformal symmetry \cite{Drummond:2007au}, and the celebrated duality
with string theory on $AdS_5 \times S^5$ \cite{Maldacena:1997re}.
There is some hope to find expressions for the amplitudes that are valid for
all values of the t'Hooft coupling, at least in the multi-color limit.\\
Such hopes were first nurtured by the intriguing BDS formula of Bern,
Dixon and Smirnov \cite{Bern:2005iz}.
It encapsulates the known infrared and collinear behavior of $n$-particle
maximally helicity violating (MHV) amplitudes in the planar approximation.
The authors of \cite{Bern:2005iz} conjectured the BDS formula to determine
the amplitudes at each loop order $L\ge2$, possibly up to some additive
finite function $R^{(n)}$ of the kinematic variables, the so-called remainder
function. Initially, $R^{(n)}$ was suspected to vanish, i.e.\ the BDS formula
was believed to be exact.\\
Both gauge and string theory arguments subsequently confirmed this
suspicion for $n=4,5$. In the weakly coupled theory, perturbative computations
uncovered the before mentioned dual conformal symmetry of scattering
amplitudes \cite{Drummond:2007au}. It implies that the remainder functions
$R^{(n)}$ can only depend on conformal cross ratios, i.e.\ on conformally
invariant combinations of the usual kinematic variables. Since there are
no such cross ratios for $n=4,5$, the corresponding remainder functions
have to be trivial. In other words, dual conformal invariance predicts
that the BDS formula is exact for $n=4,5$ to all loop orders.
This prediction was confirmed by a string theory computation of the
leading term at strong coupling \cite{Alday:2007hr}. We shall say a bit
more about the string theoretic analysis below.\\
On the other hand, the remainder function $R^{(n)}$ is now known to be non-zero for $n>5$ and
beyond one loop \cite{Drummond:2007bm,Bern:2008ap}. Several authors have described tests of the
BDS formula that exclude a vanishing remainder function. One of the most direct ways to see
that $R^{(n)} \neq 0$ is based on a study of the SYM scattering amplitudes in the
leading logarithmic approximation, see \cite{Bartels:2008ce,Bartels:2008sc}.
The high energy (Regge) limit probes the remainder function near special points in the space
of kinematic variables. While the Regge limit of the function $R^{(6)}$ vanishes at some of
these points, for example when the limit is taken with all energies negative, the authors of \cite{Bartels:2008ce,Bartels:2008sc} were able to identify one region in which the Regge
limit of $R^{(6)}$ is non-zero. Hence, $R^{(6)}$ must be a non-vanishing function of the
kinematic variables. The analysis shows how computations in the Regge limit can provide
strong and highly efficient constraints on the remainder function and its analytical
structure.\\
In the meantime, the analytic expression \cite{Goncharov:2010jf} for the exact two-loop
calculation of the six-point function \cite{DelDuca:2009au,DelDuca:2010zg} was used to
perform the relevant analytic continuation into the region with non-vanishing Regge
limit \cite{Lipatov:2010qg}. The results are in full agreement with \cite{Bartels:2008sc}.
This settles the remainder function in the two-loop approximation, and it supports
the all-order leading log generalization in \cite{Bartels:2008ce,Bartels:2008sc}.
More recently, progress has been made with the extension of the calculation of $R_6^{(n)}$ to $n>2$: in \cite{Dixon:2011pw} the symbol of $R_6^{(3)}$ has been determined (up to two parameters), and in \cite{CaronHuot:2011kk} this result has been confirmed, fixing also the two previously unknown constants.
In \cite{Dixon:2012yy} the authors quote results for the symbols of $R_6^{(n)}$ for four loops
(again up to a number of unknown constants) \cite{DDP}.
The form of the scattering amplitudes in the Regge limit which, at weak coupling, was derived in the leading logarithmic approximation is quite general, and is expected to be valid also outside the weak coupling limit.  As a function of the energy variables the amplitude contains Regge cut terms with power like dependence on s-like kinematic invariants (see below).
The exponents depend on the kinematical region and are determined by the lowest eigenvalue of the BFKL color-octet Hamiltonian for an $n$-gluon system.
These eigenvalues have recently been calculated in NLO accuracy in \cite{Fadin:2011we}, and in \cite{Dixon:2012yy} in next-to-next-to-leading order (NNLO). The power-like energy dependence of the scattering amplitudes is multiplied by Regge impact factors which are now known also in NLO \cite{Lipatov:2010ad} and even in N$^3$LO accuracy \cite{Dixon:2012yy}. A first generalization of the leading logarithmic analysis to the 7-point amplitude has been started in \cite{Bartels:2011ge}.\\
The BFKL color-octet Hamiltonian possesses the very interesting property that it coincides with the Hamiltonian of an integrable open spin chain \cite{Lipatov:2009nt} in leading order. Hence, the weakly
coupled theory provides direct evidence for integrability in the high
energy behaviour of planar scattering amplitudes.\\
Having reviewed all these results from gauge theory it is natural to
ask what string theory has to say about the high energy limit of the
remainder function $R^{(n)}$. In order to understand how the issue
can be addressed, we need to briefly sketch the development that was
initiated by the work \cite{Alday:2007hr} of Alday and Maldacena.
The main insight of this paper was the identification of the leading
contribution to an n-gluon amplitude at strong coupling with the area
$A_n$ of some 2-dimensional surface $S_n$ inside $AdS_5$.
According to the prescription of \cite{Alday:2007hr}, $S_n$ ends on a
piecewise light-like polygon on the boundary of $AdS_5$.
The light-like segments of this polygon are given by the momenta $p_j$
of the external gluons.
For $n=4$ it is possible to find the surface explicitly and the resulting
amplitude matches the prediction of the BDS formula.
Constructing $S_n$ for $n> 5$, however, turned out to be a rather difficult
problem, at least for finite $n$ and generic choice of the external momenta.
The issue was resolved through a series of papers
\cite{Alday:2009yn,Alday:2009dv,Alday:2010vh} in which the area of $S_n$ is
related to the free energy of some auxiliary quantum integrable system.
More precisely, it was argued that $A_n$ may be computed from a family of
functions $\Yf_{a,s}$ with $s = 1,2,3$ and $a = 1, \dots ,n-5$.
The latter can be determined by solving a set of coupled non-linear integral
equations. Very similar mathematical structures are familiar from the study of
{\em ground states} in 1-dimensional quantum integrable systems on a circle of
finite radius $R$. Moreover, the functional $A_n$ resembles expressions
for the free energy of such systems. So, in the sense we described, Alday
et al.\ designed a 1-dimensional quantum integrable system such that its
free energy computes the value of the remainder function $R^{(n)}$ at
strong coupling. In the 1-dimensional theory one can tune $n-5$ complex
mass parameters and the same number of real chemical potentials. The
dependence of the free energy on these parameters captures the
dependence of $R^{(n)}$ on the relevant kinematic variables. \\
Within the 1-dimensional quantum system it is natural to consider a limit
in which the masses are sent to infinity or, equivalently, the volume $R$
of the 1-dimensional space becomes large. In such a limit, all computations
simplify once finite size corrections can be neglected. This applies in
particular to the free energy of the ground state. The main goal of our
work is to show that such a large volume limit of the 1-dimensional
system possesses a nice re-interpretation in terms of the 4-dimensional
gauge theory: It corresponds to the multi-Regge limit. Put differently,
the map between 4-dimensional kinematic variables and parameters of the
1-dimensional system sends the multi-Regge regime to a point at which
all the mass parameters become large. A more precise formulation of the
limit in the 4-dimensional gauge theory will be given in section
\ref{sec:kin}. The identification \eqref{eq:mrlphi} of the corresponding
regime in the auxiliary quantum system is one of the main results of this
work. It is derived in sections \ref{sec:ew},\ref{sec:npt} and generalizes
previous observations \cite{Bartels:2010ej} for the case of six gluons to
an arbitrary number of external particles.\\
If we were only interested in the ground state energy of the system, the
large mass limit would be of limited interest. But it turns out that some
excited states of the 1-dimensional quantum system also play an important
role. In order to see them enter let us recall that the Regge limit of scattering
amplitudes can be taken in different regions of the kinematic variables, such
as the Euclidean region, the physical region where all energies are positive or
`mixed' physical regions with positive and negative energies. The limiting
value of the remainder function depends on the region. In fact, when we pass
from one region into another by continuation in the kinematic variables, the
amplitude picks up Regge cut contributions that may have a non-vanishing high
energy limit. In this sense, values of the remainder functions in the
multi-Regge limit of different kinematic regions probe the analytical structure
of the amplitude. One may wonder what all this corresponds to within the
1-dimensional auxiliary system. Since the kinematic variables are mapped
to system parameters (masses and chemical potentials), we must vary the
latter in order to move from one region of the kinematic variables to
another. In the 1-dimensional system such a variation of system parameters
can lead to a pair-wise creation of excitations above the ground state
\cite{Dorey:1996re,Dorey:1997rb}. The energy of such excited states may
be non-zero in the large volume limit. Since the energy in the 1-dimensional
system is related to the remainder function, {\em excitations} of the auxiliary
model correspond to Regge cut contributions in the gauge theory. One example of
this phenomenon was worked out in \cite{Bartels:2010ej} for the case of
six external gluons.\\
Combining the insights from the previous two paragraphs we must address the
challenge of computing excitation energies in the infinite volume limit. When
finite volume corrections can be neglected, excitation energies are determined 
by a set of algebraic Bethe ansatz equations. These replace the more complicated
non-linear integral equations that govern a 1-dimensional integrable system
at finite volume. The data that enter the Bethe ansatz equations, namely the 
momenta and $2 \mapsto 2$ scattering phases, can be derived from the non-linear 
integral equations. We will explain the general construction in section \ref{sec:ba}. 
In the case of six external gluons the derivation of the relevant Bethe ansatz is
particularly simple so that we can make things very explicit. Starting from
$n=7$, an interesting new feature appears. In going to the multi-Regge regime
of the 1-dimensional quantum system experiences wall-crossing, i.e.
the associated non-linear integral equations pick up additional
terms which we will compute in section \ref{sec:ba}. One can perform the large
volume limit of such modified integral equations, but that leads to
modifications in the Bethe ansatz, as well. More explanations and
explicit formulas are included in section \ref{sec:ba} along with a sketch of
how one may proceed to bring the Bethe ansatz equations for $n
\geq 7$ into the standard form.\\
From the point of the auxiliary quantum integrable system, the multi-Regge limit
is opposite to the high-temperature (small mass, $R$) limit of the $\Yf$-system
that was considered by Alday et.\ al.\ \cite{Alday:2010vh} and then studied in more
detail in \cite{Hatsuda:2010vr,Hatsuda:2011jn,Hatsuda:2011ke}. In terms of the
4-dimensional kinematics, the high-temperature limit corresponds to the case
where the gluon momenta $p_i$ form a regular polygon that can be embedded in
a subspace $\mathbb{R}^{1,1}$ of the full momentum space $\mathbb{R}^{1,3}$.
Another limiting regime of the kinematic variables is probed by the operator
product expansions (OPE) of polygonal Wilson loops, see \cite{Alday:2010ku,%
Gaiotto:2010fk,Gaiotto:2011dt}. The information encoded in such Wilson loop
OPEs seems more closely related to the multi-Regge limit, though the precise
link is a bit difficult to establish even at weak coupling \cite{Bartels:2011xy}.

\section{Multi-Regge kinematics}
\label{sec:kin}

In this section we discuss the relevant variables and kinematics
necessary for the description of $2\rightarrow n-2$ scattering
in the multi-Regge limit. The multi-Regge limit is characterized
by the behaviour of a particular set of Mandelstam invariants.
The remainder function of scattering amplitudes in \N4 SYM theory,
on the other hand, depends only on very special cross ratios of
Mandelstam variables which are invariant under dual conformal
symmetry. Our task here is to describe the multi-Regge limit in
terms of such cross ratios.

\subsection{Kinematic variables}

\begin{figure}[htb]
\centering
\epsfig{file=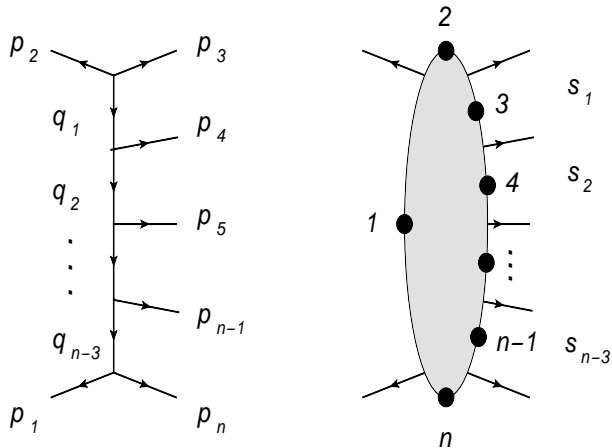, width=8cm,height=6cm}
\caption{Kinematics of the scattering process $2 \to n-2$. On the right-hand side we show a graphical representation of the dual variables $x_i$.}
\label{fig:configuration}
\end{figure}

\begin{figure}
\centering
\epsfig{file=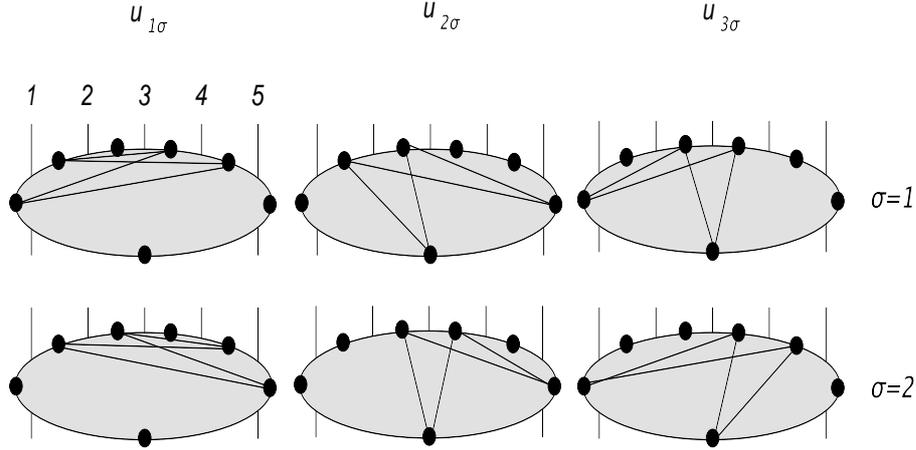, width=12cm,height=6cm}
\caption{Graphical representation for the cross ratios of the 7-point amplitude.}
\label{fig:cr}
\end{figure}

We are interested in the scattering of two incoming particles with
momenta $-p_1$, $-p_2$ resulting in a $(n-2)$-particle final state with
outgoing momenta $p_3,...,p_{n}$ as shown in figure
\ref{fig:configuration}. It will be convenient to label momenta
$p_i$ by arbitrary integers $i$ such that $p_{i+n}=p_i$.
\\
In the context of \N4 SYM theory it is advantageous to pass to
a set of dual variables $x_i$ such that
\begin{equation}
p_i=x_{i-1}-x_i.
\end{equation}
The variables $x_i$ inherit their periodicity $x_{i+n}=x_i$ from
the periodicity of the $p_i$ and momentum conservation. Let us also
introduce the notation $x_{ij} = x_i - x_j$. The $x_{ij}$ provide a
large set of Lorentz invariants $x_{ij}^2 = x_{ji}^2$.\footnote{
Throughout this paper we use the metric $(1,-1,-1,-1)$.} When
expressed in terms of the momenta, these read
\begin{equation} \label{eq:xij2}
x^2_{ij}=\left(p_{i+1}+\cdots+p_j\right)^2.
\end{equation}
Obviously, only $3n-10$ Lorentz invariants are independent.
Throughout this text we shall describe scattering processes
through the $3n-10$ independent variables
\begin{eqnarray}
t_r & = & x^2_{1,r+2}\ ,
\label{eq:def_t}\\[4mm]
s_r & = & x^2_{r+1,r+3}\ ,
\label{eq:def_s_r}\\[2mm]
\eta_p & =& \frac{x^2_{p-2,p} x^2_{p-1,p+1}}{x^2_{p-2,p+1}}
\label{eq:def_eta}\ ,
\end{eqnarray}
where $r=1,...,n-3$ extends over all $t$-channels and
$p=4,...,n-1$ labels the produced particles. In discussions
of the multi-Regge limit it is actually quite common to use the
cosines $z_r = \cos \theta_r$ of the scattering angle $\theta_r$
defined in the CM-system of the momentum $q_r$ instead of $s_r$
and to replace our variables $\eta_p$ by the so-called Toller
angles $\omega_{p\,p+1}$. With these choices, the multi-Regge
limit is obtained by sending $z_r \to \infty$ with $t_r$ and
$\omega_{p\,p+1}$ held fixed. The variables we defined in eqs.\
(\ref{eq:def_s_r}) and (\ref{eq:def_eta}) are more convenient
(for a pedagogical discussion see \cite{Brower:1974yv}). In the
variables \eqref{eq:def_t}-\eqref{eq:def_eta} the multi-Regge
limit is taken by sending $s_r$ to infinity while keeping both
$t_r$ and $\eta_p$ fixed.\\
As we recalled in the introduction, the missing remainder
functions for gluon scattering in \N4 SYM theory possess dual
conformal symmetry. In other words, they only depend on
conformally invariant combinations of Mandelstam variables.
For an n-gluon scattering amplitude there are only
$3n-15$ independent conformal invariants. We choose to
work with the following set of cross ratios:
\begin{align}
u_{1\sigma}&=\frac{x^2_{\sigma+1,\sigma+5}x^2_{\sigma+2,\sigma+4}}
{x^2_{\sigma+2,\sigma+5}x^2_{\sigma+1,\sigma+4}},\label{eq:cr1}\\
u_{2\sigma}&=\frac{x^2_{\sigma+3,n}x^2_{1,\sigma+2}}
{x^2_{\sigma+2,n}x^2_{1,\sigma+3}},\label{eq:cr2}\\
u_{3\sigma}&=\frac{x^2_{2,\sigma+3}x^2_{1,\sigma+4}}
{x^2_{2,\sigma+4}x^2_{1,\sigma+3}},
\label{eq:cr3}
\end{align}
where $\sigma=1,...,n-5$. For the $2\to5$ scattering process,
a convenient graphical representation of the cross ratios is
displayed in figure \ref{fig:cr}. Our main task now is to analyze
the behavior of these $3n-15$ cross ratios in the multi-Regge
limit.

\subsection{Scattering in the center-of-mass system}
\label{sec:q_kin}

To study the behaviour of the cross ratios in the multi-Regge limit
we need some preliminary results which we will obtain by specializing
to the CM-frame, writing the results in Lorentz invariant form. In
our analysis throughout this subsection we shall study the behaviour
of the set of subenergies
\begin{equation}
 s_{i\cdots j}=(p_i+\cdots +p_j)^2\ .  \label{eq:subenergies}
\end{equation}
Note that these subenergies are the same as the Lorentz
invariant variables $x_{ij}^2=s_{i+1 \cdots j}$ we introduced in
the previous subsection. The only reason we change notation here
is to give the equations in this section a more familiar
form. Subenergies $s_{i i+1}$ for two adjacent particles
of momenta $p_i$ and $p_{i+1}$ are related to our variables $s_r$
through $s_{r} = s_{r+2\,r+3}$. Furthermore, the total energy $s =
(p_1+p_2)^2$ of the process is given by $s = s_{3\cdots n}$.\\
In studying the Regge behaviour of the subenergies, it is useful
to introduce the Sudakov parametrization
\begin{equation}
q_r=\delta_r \hat{p}_1+\gamma_r \hat{p}_2 + q_{r\perp},
\end{equation}
for $r = 1, \dots, n-3$. Here, $\hat{p}_1$ and $\hat{p}_2$ are light-like reference vectors from which we define our incoming momenta as $p_1=-\hat{p}_1$, $p_2=-\hat{p}_2$.
They obey $2\hat{p}_1\hat{p}_2=2p_1p_2=s$, and the transverse part, $q_{r\perp}$, is orthogonal
to both $\hat{p}_1$ and $\hat{p}_2$,  i.e.\ $\hat{p}_1q_{r\perp} = \hat{p}_2q_{r\perp} =0$. A
convenient frame is the CM-system of the incoming particles 1 and 2,
with momenta $\hat{p}_1$ and $\hat{p}_2$ along the z-direction. We can determine
the Sudakov parameters $\gamma_r$ and $\delta_r$ by considering the
following subenergies,
\begin{align}
s_{3\cdots r+2}&=(p_3+\cdots+p_{r+2})^2=(-p_2-q_1+q_1-q_2+\cdots+
q_{r-1}-q_{r})^2=(p_2+q_{r})^2\nonumber\\[2mm]
s_{r+3\cdots n}&= (p_{r+3}+\cdots+p_n)^2 = (q_{r}-q_{r+1} + q_{r+1} -
\cdots + q_{n-3}-p_1)^2 =
(p_1-q_{r})^2\ . \nonumber
\end{align}
Using that $q_i^2=t_i$ we find
\begin{equation}
s_{3\cdots r+2}=2q_rp_2+t_r=-\delta_rs+t_r \quad \Rightarrow \quad
\delta_r= \frac{t_r-s_{3\cdots r+2}}{s},
\label{eq:delta}
\end{equation}
as well as
\begin{equation}
s_{r+3\cdots n}=t_r-2q_rp_1 \quad \Rightarrow
\quad
\gamma_r=\frac{s_{r+3\cdots n}-t_r}{s}.
\label{eq:gamma}
\end{equation}
Up to now all the identities have been exact. Now we would like to
continue considering the multi-Regge limit which, as defined in the
previous subsection, amounts to sending all pairwise energies $s_1,
\dots  ,s_{n-3}$ to infinity, while keeping both $t$- and $\eta$-variables
fixed. In this paper we will restrict ourselves to the physical kinematic
region where all energies are positive and all $t_r$ negative (in a future
study we we will consider also analytic continuations into other `mixed'
physical regions where some energies are negative). For the subenergies
introduced in eq.\ (\ref{eq:subenergies}) the multi-Regge limit implies
\begin{equation}
s\gg s_{3\cdots n-1}, s_{4\cdots n}\gg s_{3\cdots n-2},...,
s_{5\cdots n}\gg\cdots\gg s_1,...,s_{n-3}\gg -t_1,...,-t_{n-3}.
\label{eq:def_mrl}
\end{equation}
From this well-known hierarchy of energy variables along with eqs.\ (\ref{eq:delta}) and (\ref{eq:gamma}) one deduces the
strong ordering of the Sudakov parameters $\gamma_i$ and
$\delta_i$
\begin{equation}
1\gg \gamma_1\gg \gamma_2\gg\cdots\gg\gamma_{n-3}
\label{eq:gammaordering}
\end{equation}
and
\begin{equation}
1\gg-\delta_{n-3}\gg-\delta_{n-4}\gg\cdots\gg-\delta_1.
\label{eq:deltaordering}
\end{equation}
As a simple consequence of eqs.\ (\ref{eq:delta}) and (\ref{eq:gamma})
we note that in the multi-Regge limit
\begin{equation}
t_r = q_r^2 = s \gamma_r \delta_r + q_{r\perp}^2 \cong  q_{r\perp}^2,
\end{equation}
where we could drop the term $s \gamma_r \delta_r \cong - s^{-1}
s_{3\cdots r+2} s_{r+3\cdots n}$ because of the strong ordering
\eqref{eq:def_mrl} of subenergies in the multi-Regge limit. In
conclusion, the finiteness of the $t_r$ in the multi-Regge limit
implies that the transverse components of the $q_r$ stay finite.
Since $p_{r+3}= q_{r}-q_{r+1}$ this is also true for the transverse
components of the momenta $p_{r+3}$ for $r=1,\dots,n-4$. We can
compute this finite quantity from the mass-shell conditions of
the produced particles with momenta $p_4,\dots  ,p_{n-1}$:
\begin{align}
0=p_p^2=(q_{p-3}-q_{p-2})^2 \cong - s \gamma_{p-3} \delta_{p-2} +
p_{p\perp}^2 = \frac{s_{3\cdots p} s_{p\cdots n}}{s} + p_{p\perp}^2
\nonumber \\[2mm] \Rightarrow
\frac{s_{3\cdots p} s_{p\cdots n}}{s} \cong  -
p_{p\perp}^2 = \vec{p}_{p}^{\;2}.
\label{eq:mass-shell}
\end{align}
We are now prepared to look at the subenergies in the multi-Regge limit.
Let us begin with the subenergies formed by two adjacent particles. We
have expressions of the form
\begin{eqnarray}
s_r=(q_{r-1} - q_{r+1})^2 & \cong &  - s \gamma_{r-1} \delta_{r+1} +
(p_{r+2}+p_{r+3})_{\perp}^2 \nn \\[3mm] & \cong &
\frac{s_{3\cdots r+3}s_{r+2\cdots n}}{s}+
(p_{r+2}+p_{r+3})_{\perp}^2
\label{eq:subenergies1}
\end{eqnarray}
where $r$ runs over $r=2,...,n-4$. Similarly, we can determine
the leading terms in the multi-Regge limit of the subenergies for
three adjacent particles,
\begin{eqnarray}
s_{p-1\,p\,p+1} = (q_{p-4}-q_{p-1})^2 & \cong&  -s \gamma_{p-4} \delta_{p-1} +
(p_{p-1}+p_{p}+p_{p+1})_{\perp}^2 \nn\\[3mm]
& \cong&   \frac{s_{3\cdots p+1}s_{p-1\cdots n}}{s} +
(p_{p-1}+p_{p}+p_{p+1})_{\perp}^2
\label{eq:subenergies3}
\end{eqnarray}
for $p=5,...,n-2$. As an application of these results we can now
express the variables $\eta_p$ in the multi-Regge limit through the
momenta of produced particles. In order to do so, we express the
basic definition \eqref{eq:def_eta} of the $\eta$ variables through
the subenergies \eqref{eq:subenergies}. All three subenergies that
appear in the expression can then be replaced by their leading
behavior in the multi-Regge limit, i.e.\ the first term in
eqs.\ ({\ref{eq:subenergies1}) and ({\ref{eq:subenergies3}),
respectively. Comparing the resulting expression with the result
\eqref{eq:mass-shell} we arrive at
\begin{equation} \label{eq:etap}
\eta_p = \frac{s_{p-3}s_{p-2}}{s_{p-1\,p\, p+1}}
\cong -p_{p \perp}^2 = \vec{p}_{p}^{\,2}\ .
\end{equation}
We can apply this result to derive a relation between our $\eta$ variables
and the (azimuthal) Toller angles between between adjacent vectors $\vec{q}_r$,
$\vec{q}_{r+1}$. The angle between vectors $\vec{q}_r$ and $\vec{q}_{r+1}$
can be determined by computing
\begin{equation}
\vec{p}_{r+3}^{\,2}=(\vec{q}_r-\vec{q}_{r+1})^2=|t_r|+|t_{r+1}|-2\sqrt{|t_r||t_{r+1}|}
\cos\theta_{r,r+1}
\end{equation}
for $r = 1, \dots, n-4$. According to eq.\ (\ref{eq:etap}), the quantity $\vec{p}_{r+3}$
coincides with the variable $\eta_{r+3}$ in the multi-Regge limit. Hence, we obtain
\begin{equation}
\cos\theta_{r,r+1} \cong \frac{|t_r|+|t_{r+1}|-\eta_{r+3}}{2\sqrt{|t_r||t_{r+1}|}}.
\end{equation}
Below we shall need an explicit expression for the sine of the (azimuthal)
Toller angle $\theta_{r,r+1}$ in the multi-Regge limit. It is given by
\begin{equation}
\sin\theta_{r,r+1}=\sqrt{1-\cos^2\theta_{r,r+1}} \cong
\frac{\lambda(|t_r|,|t_{r+1}|,\eta_{r+3})}{2\sqrt{|t_r||t_{r+1}|}},
\end{equation}
with
\begin{equation}
\lambda^2(|t_r|,|t_{r+1}|,\eta_{r+3})=
2|t_r||t_{r+1}|+2\eta_{r+3}|t_r|+2\eta_{r+3}|t_{r+1}|-|t_r|^2-|t_{r+1}|^2
-\eta_{r+3}^2.
\end{equation}
From these angles between adjacent vectors $\vec{q}_r$ and $\vec{q}_{r+1}$ it
is straightforward to compute the angles between arbitrary vectors $\vec{q}_r$
and $\vec{q}_{r'}$ due to the 2-dimensional kinematics in the multi-Regge limit.\\
So far, we have only looked at subenergies for up to three particles. But it
is clear how to continue the analysis. Generalizing the derivations of our
eqs.\ (\ref{eq:subenergies1}) or (\ref{eq:subenergies3}) we obtain
\begin{equation}
s_{r+2 \dots r'+3} \cong \frac{s_{r+2\cdots n} s_{3\cdots r'+3}}{s} +
(p_{r+2} + \dots p_{r'+3})^2_\perp \ . \label{eq:subenergies4a}
\end{equation}
Comparison of the leading terms allow us to conclude that
\begin{equation}
s_{r+2\cdots r'+3}=(p_{r+2} + \cdots + p_{r'+3})^2 \cong
\frac{s_r\cdots s_{r'}}{\eta_{r+3}\cdots\eta_{r'+2}}\ .
\label{eq:subenergies4}
\end{equation}
Note that for $r'=r+1$, i.e.\ when the subenergy on the left hand
side involves three particles, the relation is exact and not restricted
to the multi-Regge limit. For more than three particles contributing to
the subenergy, on the other hand, the result \eqref{eq:subenergies4}
only describes the leading term and it takes more effort to determine
the subleading term from eq.\ (\ref{eq:subenergies4a}). In the next
subsection we only need to determine the subleading contribution
for very special combinations of subenergies. We postpone further
discussion of such subleading terms until we have spelled out the
relevant combinations.

\subsection{Cross ratios in the multi-Regge limit}

Let us now look at the behavior of the cross ratios defined
by eqs.\ (\ref{eq:cr1})-(\ref{eq:cr3}) in the multi-Regge
limit. Combining eq.\ (\ref{eq:subenergies4}) with eq.\
(\ref{eq:xij2}) we obtain the leading term in the multi-Regge
limit of the basic Lorentz invariants $x_{1+r,3+r'}^2$:
\begin{align}
 x_{1+r,3+r'}^2 = s_{r+2\cdots r'+3}
\cong \frac{s_r\cdots s_{r'}}{\eta_{r+3}\cdots\eta_{r'+2}} \ .
\label{eq:subenergies5}
\end{align}
If we insert this asymptotic behaviour into our definitions
(\ref{eq:cr2}) and (\ref{eq:cr3}) for the cross ratios
$u_{2\sigma}$ and $u_{3\sigma}$ we conclude
\begin{align}
u_{2\sigma}&=\frac{t_\sigma}{t_{\sigma+1}}
\frac{x^2_{\sigma+3,n}}{x^2_{\sigma+2,n}}\cong
\frac{t_\sigma}{t_{\sigma+1}}\frac{\eta_{\sigma+4}}
{s_{\sigma+1}} ,\label{eq:mrlu2} \\[2mm]
u_{3\sigma}&=\frac{t_{\sigma+2}}{t_{\sigma+1}}
\frac{x^2_{2,\sigma+3}}{x^2_{2,\sigma+4}}
\cong\frac{t_{\sigma+2}}{t_{\sigma+1}}
\frac{\eta_{\sigma+3}}{s_{\sigma+1}} \ .
\label{eq:mrlu3} \end{align}
When we send $s_\sigma$ to infinity to reach the multi-Regge
regime, both sets of cross ratios go to zero. Since the coefficients
of $1/s_\sigma$ depend on the $t-$ and $\eta$ variables only, the
ratios
\begin{equation} \label{eq:ratiou2u3}
\frac{u_{2\sigma}}{u_{3\sigma}} \cong \frac{t_{\sigma}}{t_{\sigma+2}}
\frac{\eta_{\sigma+4}}{\eta_{\sigma+3}}
\end{equation}
approach a non-vanishing constant value in the multi-Regge limit.
Functions of the cross ratios that remain finite in the multi-Regge
limit can therefore depend on the ratios $u_{2\sigma}/u_{3\sigma}$.\\
Let us now look at the remaining set of cross ratios $u_{1\sigma}$.
Once more, we can rewrite our definition \eqref{eq:cr1} in terms of
the energy variables \eqref{eq:subenergies} using eq.\ (\ref{eq:xij2})
to obtain
\begin{equation}
u_{1\sigma} =
\frac{s_{\sigma+2\cdots\sigma+5} s_{\sigma+3\,\, \sigma+4}}
{s_{\sigma+3\cdots\sigma+5} s_{\sigma+2 \cdots\sigma+4}}\ .
\end{equation}
If we now insert the limiting behavior \eqref{eq:subenergies4} for all
four subenergies we see that all variables $u_{1\sigma}$ behave as
$u_{1\sigma}\cong 1+ \dots$. In order to obtain the leading non-trivial
term in the multi-Regge limit we must work a little harder.\\
To this end we write, in analogy with eqs.\
(\ref{eq:subenergies1})-(\ref{eq:subenergies3}), expressions for the subenergies
$s_{\sigma+3\,\, \sigma+4}$, $s_{\sigma+3\cdots\sigma+5}$, $s_{\sigma+2\cdots\sigma+4}$, and $s_{\sigma+2\cdots\sigma+5}$. Beginning with the relation for
$s_{\sigma+3\,\, \sigma+4}$
\begin{equation}
1=\frac{s_{3\cdots\sigma+4} s_{\sigma+3 \cdots n}}
{s s_{\sigma+3\,\, \sigma+4}} + \frac{(p_{\sigma+3} + p_{\sigma+4})_{\perp}^{\,2}}{s_{\sigma+3\,\, \sigma+4}},
\end{equation}
we write
\begin{align}
1 = \frac{s_{3\cdots\sigma+4} s_{\sigma+2 \cdots n}}{s s_{\sigma+2\cdots \sigma+4}}
\cdot
\frac{s_{3\cdots \sigma+5} s_{\sigma+3 \cdots n}}{s s_{\sigma+3\cdots \sigma+5}}
\cdot
\frac{s s_{\sigma+2 \cdots \sigma+5}} {s_{3\cdots \sigma+5} s_{\sigma+2\cdots n}}
\cdot
\frac{s_{\sigma+2\cdots \sigma+4} s_{\sigma+3 \cdots \sigma+5}}
{s_{\sigma+3\,\, \sigma+4} s_{\sigma+2\cdots \sigma+5}}  \nonumber \\
+ \frac{(p_{\sigma+3} + p_{\sigma+4})_{\perp}^{\,2}}{s_{\sigma+3\,\, \sigma+4}}.
\label{eq:u1sigmacorrections}
\end{align}
On the right-hand side, the fourth fraction equals $u_{1\sigma}^{-1}$, and the first three fractions are equal to unity, arising from the
equations for $s_{\sigma+2\cdots \sigma+4}$, $s_{\sigma+3\cdots \sigma+5}$, and
$s_{\sigma+2\cdots \sigma+5}$,
with corrections of the order ${\cal O}(1/s_{\sigma+2\cdots \sigma+4})$,
${\cal O}(1/s_{\sigma+3\cdots \sigma+5})$, ${\cal O}(1/s_{\sigma+2\cdots \sigma+5})$, respectively.
Compared to the last term in eq.\ (\ref{eq:u1sigmacorrections}), these corrections  can be neglected and we are left with
\begin{equation}
u_{1\sigma} \cong 1 + \frac{(p_{\sigma+3} + p_{\sigma+4})_{\perp}^{\,2}}{s_{\sigma+3\,\, \sigma+4}}
\label{eq:subleading}
\end{equation}
with further corrections being smaller than ${\cal O}(1/s_{\sigma+3\,\sigma+4})$.
The numerator of the correction term in eq.\ (\ref{eq:subleading}) can be expressed in terms of Lorentz invariants. To do so, we use the results of section \ref{sec:q_kin} by writing
\begin{equation}
(\vec{p}_{\sigma+3}+\vec{p}_{\sigma+4})^2=(\vec{q}_{\sigma}-\vec{q}_{\sigma+2})^2=|t_{\sigma}|+|t_{\sigma+2}|-2\sqrt{|t_{\sigma}||t_{\sigma+2}|}\cos(\theta_{\sigma,\sigma+1}+\theta_{\sigma+1,\sigma+2}),
\end{equation}
which finally gives $(\vec{p}_{\sigma+3}+\vec{p}_{\sigma+4})^2 = \rho_\sigma$ with
a set of functions of the $t$ and $\eta$ variables defined by
\begin{align}
\nonumber \rho_\sigma(t,\eta) &:=|t_{\sigma}|+|t_{\sigma+2}|-2\sqrt{|t_{\sigma}||t_{\sigma+2}|}
\left(\frac{|t_{\sigma}|+|t_{\sigma+1}|-\eta_{\sigma+3}}{2\sqrt{|t_{\sigma}||t_{\sigma+1}|}}
\frac{|t_{\sigma+1}|+|t_{\sigma+2}|-\eta_{\sigma+4}}{2\sqrt{|t_{\sigma+1}||t_{\sigma+2}|}}\right.\\[2mm]
&\hspace*{2cm} \left.-\frac{\lambda(|t_{\sigma}|,|t_{\sigma+1}|,\eta_{\sigma+3})}{2\sqrt{|t_{\sigma}|
|t_{\sigma+1}|}}\frac{\lambda(|t_{\sigma+1}|,|t_{\sigma+2}|,\eta_{\sigma+4})}
{2\sqrt{|t_{\sigma+1}||t_{\sigma+2}|}}\right)\ .
\end{align}
We can now summarize the findings of our analysis on the multi-Regge limit of
the cross ratios $u_{1\sigma}$ through
\begin{equation}
u_{1 \sigma} - 1 = \rho_{\sigma}(t,\eta)/s_{\sigma+1} \
\label{eq:mrlu1}
\end{equation}
where we also changed notations back using $s_{\sigma+1} = s_{\sigma+3\,\sigma+4}$, as mentioned before. As in the case of the cross ratios $u_{2\sigma}$ and
$u_{3\sigma}$ the leading correction to $u_{1\sigma}-1$ vanishes in the
multi-Regge limit. But the following ratios remain finite
\begin{align}
\frac{u_{1\sigma}-1}{u_{2\sigma}} \cong  \frac{\rho_\sigma(t,\eta) t_{\sigma+1}}
{t_\sigma \eta_{\sigma+4}},  \\[2mm]
\frac{u_{1\sigma}-1}{u_{3\sigma}} \cong \frac{\rho_\sigma(t,\eta) t_{\sigma+1}}
{t_{\sigma+2} \eta_{\sigma+3}}.
\end{align}
This concludes our description of the kinematics in the multi-Regge
limit.

\section{The n-gluon thermodynamic bubble ansatz}
\label{sec:ngltherm}
The main goal of this section is to review the \Yf-system for the
computation of n-gluon amplitudes at strong coupling \cite{Alday:2009dv,
Alday:2010ku}. In the first subsection we explain how the most interesting
contribution to the scattering amplitude can be computed by solving a
system of non-linear integral equations (NLIE). Then we relate the
parameters of the NLIE to the cross ratios that were introduced in
eqs.\ (\ref{eq:cr1})-(\ref{eq:cr3}).

\subsection{Amplitudes and the \Yf-system}
\label{sec:ampy}
We are interested in the calculation of scattering amplitudes
in \N4 SYM at strong coupling. To leading order they are given by
\begin{equation}
\mathrm{Amplitude}\sim e^{-\frac{\sqrt{\lambda}}{2\pi}A},
\end{equation}
where $A$ is the area of a minimal surface in $AdS_5$ with
piece-wise light-like boundary. A general prescription for the
calculation of this area $A$ is given in \cite{Alday:2010vh,Yang2011}.
It contains a number of different pieces, including a divergent
BDS-like term and a number of finite contributions. All but one
of these terms can be spelled out explicitly. The remaining one
is also known, but it is characterized somewhat indirectly
through the solution of a coupled system of non-linear integral
equations. Because of the resemblance with the way one describes
the free energy of a 2-dimensional quantum integrable system,
this contribution to the area $A$ has been dubbed $A_{\mathrm
{free}}$. In our analysis of the multi-Regge limit we can
restrict to the discussion of this free energy contribution
since the remaining terms are straightforward to include.\\
For our study of scattering amplitudes in the multi-Regge regime
we need some more background on $A_\mathrm{free}$. It can be
calculated from a set of functions $\Yf_{a,s}$ with $a=1,2,3$ and
$s=1, \dots, n-5$, which are determined as solutions of the
following set of integral equations:
\begin{align}
\mathrm{log}\Yf_{1,s}&=-m_s\cosh\theta-C_s-
\frac{1}{2}K_2\star\beta_s- K_1\star\alpha_s-
\frac{1}{2}K_3\star\gamma_s,\label{eq:ysys1}\\[2mm]
\mathrm{log}\Yf_{2,s}&=-m_s\sqrt{2}\cosh\theta-K_2
\star\alpha_s-K_1\star\beta_s,\label{eq:ysys2}\\[2mm]
\mathrm{log}\Yf_{3,s}&=-m_s\cosh\theta+C_s-\frac{1}{2}
K_2\star\beta_s- K_1\star\alpha_s+
\frac{1}{2}K_3\star\gamma_s,\label{eq:ysys3}
\end{align}
where $K\star f$ denotes the convolution integral
\begin{align}
\int\limits_{-\infty}^{\infty} d\theta' K(\theta-\theta')f(\theta')
\end{align}
and $\alpha_s$, $\beta_s$, $\gamma_s$ are given by
\begin{align}
\alpha_s&=\mathrm{log}\frac{(1+\Yf_{1,s})(1+\Yf_{3,s})}
{(1+\Yf_{2,s-1})(1+\Yf_{2,s+1})},\label{eq:alpha}\\[2mm]
\beta_s&=\mathrm{log}\frac{(1+\Yf_{2,s})^2}{(1+\Yf_{1,s-1})
(1+\Yf_{1,s+1})(1+\Yf_{3,s-1})(1+\Yf_{3,s+1})},\label{eq:beta}\\[2mm]
\gamma_s&=\mathrm{log}\frac{(1+\Yf_{1,s-1})
(1+\Yf_{3,s+1})}{(1+\Yf_{1,s+1})(1+\Yf_{3,s-1})}. \label{eq:gammaY}
\end{align}
The kernel function $K_a$ are known to take the form
\begin{align} \label{eq:kernels}
K_1=\frac{1}{2\pi}\frac{1}{\cosh\theta} \quad , \quad
K_2=\frac{\sqrt{2}}{\pi}\frac{\cosh\theta}{\cosh 2\theta} \quad
, \quad K_3=\frac{i}{\pi}\tanh 2\theta.
\end{align}
Furthermore, $m_s$ and $C_s$ are constants that we need to determine in the following.
These equations can be used for $\left|\mathrm{Im}\,\theta\right|\leq\frac{\pi}{4}$.
For larger values of $\theta$, we can either pick up pole contributions from the
kernels or use the recursion relation
\begin{equation} \label{eq:recursion}
\Y{a}{s}{r}=\frac{\left(1+\Y{a}{s+1}{r\pm 1}\right)\left(1+\Y{4-a}{s-1}{r\pm 1}\right)}{\Y{4-a}{s}{r\pm 2}\left(1+\frac{1}{\Y{a+1}{s}{r\pm 1}}\right)\left(1+\frac{1}{\Y{a-1}{s}{r\pm 1}}\right)},
\end{equation}
where we introduced the symbol $\Y{a}{s}{r}(\theta) = \Yf_{a,s}(\theta + i r \pi/4)$ for
$\Yf$-functions with arguments shifted by multiples of $i \pi /4$. For the moment let us
consider the $m_s$ as complex parameters while we take $C_s$ to be real. Consequently,
the total number of real parameters in the eqs.\ \eqref{eq:ysys1}-\eqref{eq:ysys3}
is $3(n-5)$, matching the number of independent cross ratios for $n$-gluon scattering.
The precise relation between the cross ratios and the parameters $m_s$, $C_s$ will be
addressed in the next subsection. Right now it suffices to keep in mind that the
parameters $m_s = m_s(u_{a\sigma})$ and $C_s = C_s(u_{a\sigma})$ in the $\Yf$-system
need to be adjusted as we vary the kinematic variables.\\
Let us add a few more comments on complex mass parameters $m_s$. The correct way
to interpret the $\Yf$-system in the presence of complex masses is through the
following substitutions in the original equations:
\begin{equation}
m_s\rightarrow\left|m_s\right|,\quad \Yf_{a,s}(\theta)\rightarrow \Yf_{a,s}(\theta+i\phi_s),\quad K^{a,a'}_{s,s'}(\theta-\theta')\rightarrow K^{a,a'}_{s,s'}(\theta-\theta'+i(\phi_s-\phi_{s'})).
\end{equation}
Here we have split each complex $m_s$ into the real parameters $|m_s|$ and the
phase $\phi_s$. For $\left|\phi_s-\phi_{s'}\right|=n\cdot\frac{\pi}{4}$ the kernels
become singular, and we have to pick up the corresponding poles. Such large phases
are going to play an important role later on.\\
Once the solution $Y_{a,s}$ of the $\Yf$ system has been found, we can compute
the quantity $A_\mathrm{free}$ through the simple prescription
\begin{equation} \label{eq:Afree}
A_\mathrm{free}=\sum\limits_s\int \frac{d\theta }{2\pi}
\left|m_s\right|\cosh\theta\mathrm{log}
\left[\left(1+\Yf_{1,s}\right)\left(1+\Yf_{3,s}\right)
\left(1+\Yf_{2,s}\right)^{\sqrt{2}}\right]\left(\theta+i\phi_s\right).
\end{equation}
Note that $A_\mathrm{free}$ depends on the kinematic variables $u_{a\sigma}$
of the scattering process through the parameters $m_s$, $\phi_s$ and $C_s$ of
the auxiliary quantum integrable system.

\subsection{\Yf-system and cross ratios}
In this section we relate the cross ratios to the value of the
\Yf-functions at special values of the spectral parameter $\theta$.
To do so, we follow \cite{Alday:2010vh} and define
\begin{equation}
U_s^{[r]}:= 1+ \frac{1}{\Ynt{2}{s}{r}} :=
1+\left.\frac{1}{\Yf_{2,s}}\right|_{\theta=i\pi r/4}
= 1+\left.\frac{1}{\Y{2}{s}{r}}\right|_{\theta=0}
\end{equation}
for $s=1, \dots,n-5$ and $r$ any integer. These quantities
possess a rather simple relation with the cross ratios. When
both indices of $U$ are even, one has
\begin{equation}
U_{2k-2}^{[2p]}=\frac{x^2_{-k+p,k+p}x^2_{-k+p-1,k+p-1}}
                   {x^2_{-k+p-1,k+p}x^2_{-k+p,k+p-1}}\ .
\label{eq:gen_crev}
\end{equation}
Here $2k-2$ is the number of cusps between the pairs
$\{x_{-k+p-1},x_{-k+p}\}$ and $\{x_{k+p-1},x_{k+p}\}$, see
\cite{Alday:2010vh} for details.
For sites separated by an odd number $2k-1$ of cusps, the
relevant relation reads
\begin{equation} \label{eq:gen_crodd}
U_{2k-1}^{[2p+1]} = \frac{x^2_{-k+p,k+p+1}x^2_{-k+p-1,k+p}}
                   {x^2_{-k+p-1,k+p+1}x^2_{-k+p,k+p}}\ .
\end{equation}
We can now insert the general relations \eqref{eq:gen_crev} and
\eqref{eq:gen_crodd} into our definition of the special
cross ratios eqs.\ (\ref{eq:cr1})-(\ref{eq:cr3}) to obtain
\begin{align}
u_{1\sigma}&=\frac{x^2_{\sigma+1,\sigma+5}x^2_{\sigma+2,\sigma+4}}
 {x^2_{\sigma+2,\sigma+5}x^2_{\sigma+1,\sigma+4}} =
 \left(U_1^{[2\sigma+7]}\right)^{-1}=\U{1}{2\sigma+7}\nn \\[2mm]
u_{2\sigma}&=\frac{x^2_{\sigma+3,n}x^2_{1,\sigma+2}}{x^2_{\sigma+2,n}x^2_{1,\sigma+3}}\ \ =
\quad \left(U_\sigma^{[\sigma+4]}\right)^{-1}\ =\ \ \U{\sigma}{\sigma+4}\label{eq:spe_cr}\\[2mm]
u_{3\sigma}&=\frac{x^2_{2,\sigma+3}x^2_{1,\sigma+4}}
{x^2_{1,\sigma+3}x^2_{2,\sigma+4}}\ \ = \quad \left(U_\sigma^{[\sigma+6]}\right)^{-1}
\ =\ \ \U{\sigma}{\sigma+6}.\nn
\end{align}
The $\Yf$-functions appearing in our theory depend on the $3n-15$ parameters
$m_s, C_s, \phi_s$ in the non-linear integral equations. Hence, $\Y{a}{s}{r}$
which are just shifted $\Yf$-function evaluated at the origin of the $\theta$
plane, are functions of these parameters. The equations \eqref{eq:spe_cr}
describe the transformation between the parameters in the non-linear integral
equations and the cross ratios of the scattering process. We can invert them,
at least numerically, to determine $m_s,C_s$ and $\phi_s$ from the kinematic
invariants of the $n$-gluon system.\\
The formulae derived above for the cross ratios give rise to large values
in the upper index of the $\Yf$-functions. As we shall see later, this is
a bit of a nuisance for practical computations. It is therefore useful to
observe that there exist two symmetries which may be used to reduce the
value of the upper index. These symmetries have their origin in the
$\mathbb{Z}_4$-symmetry of the underlying Hitchin system. The first of
these symmetries reads
\begin{equation}
U_{s}^{[l]}=U_{s}^{[l\pm 2n]}\ .
\end{equation}
Note that such a symmetry must necessarily hold in order for the
identification \eqref{eq:gen_crev} and \eqref{eq:gen_crodd} with cross ratios
to be consistent with the symmetry $x_{i+n} = x_i$ of the $x$-variables.
A second useful symmetry of the quantities $U_{s}^{[l]}$ is given by
\begin{equation}
U_s^{[l]}=U_{n-4-s}^{[l\pm n]}\ .
\label{eq:symn}
\end{equation}
In this case, we must accompany the shift in the upper by a reflection
in the lower index. Once again, the corresponding symmetry of cross ratios
is easy to verify. With these symmetries, it is always possible to reduce
the absolute value of the upper index of the $\Yf$-functions to
$\left\lfloor\frac{n}{2}\right\rfloor$ or lower. In order to
achieve further reduction, one can employ the recursion
relations \eqref{eq:recursion}.


\section{Multi-Regge limit of the TBA for $n \leq 7$ gluons}
\label{sec:ew}

The multi-Regge limit was defined in section \ref{sec:kin} through the dynamical
invariants of the scattering process as a limit in which the $s$-variables
are sent to infinity while $t$- and $\eta$-variables are held fixed. We
have also analyzed how the special cross ratios \eqref{eq:cr1}-\eqref{eq:cr3}
behave in the limit. In section \ref{sec:ngltherm} we then went on to discuss the relation
\eqref{eq:spe_cr} between cross ratios and the parameters of the non-linear
integral equations. Our next task is to understand which limit of the
parameters $m_s, \phi_s$ and $C_s$ has to be taken in order for the
cross ratios to show multi-Regge behavior. The case with $n=6$ has
been treated before \cite{Bartels:2010ej} and is relatively simple to
analyze. We will review some formulas in the next subsection before
turning to $n>6$ gluons. Beyond $n=6$ there are some important new
features in taking the multi-Regge limit. We will explain these first
for the example of $n=7$ before we delve into a general analysis in
the subsequent section.

\subsection{Review of the hexagon $n=6$}

In order to understand the basic steps of our analysis we would like
to review briefly how things work in the case of $6$ gluons \cite{Bartels:2010ej}.
For six points, there are three independent cross ratios:
\begin{flalign*}
&u_1=\frac{x^2_{35}x^2_{26}}{x^2_{36}x^2_{25}}=\frac{Y_2^{[-3]}}{1+Y_2^{[-3]}},%
&&u_2=\frac{x^2_{13}x^2_{46}}{x^2_{14}x^2_{36}}=\frac{Y_2^{[-1]}}{1+Y_2^{[-1]}},%
&&u_3=\frac{x^2_{15}x^2_{24}}{x^2_{14}x^2_{25}}=\frac{Y_2^{[1]}}{1+Y_2^{[1]}}.
\end{flalign*}
We have omitted all the second indices for both $u$ and $Y$ because $\sigma = 1$
is the only value it can take when $n-5=1$. From the kinematic analysis in
section \ref{sec:kin} we know that $u_1\rightarrow 1$ in the multi-Regge limit,
while $u_2$ and $u_3$ tend to zero. Comparing the expressions for the \Yf-functions
with this limit, the authors of \cite{Bartels:2010ej} show that
\begin{equation}
m\ \rightarrow\ \infty, \quad \phi\ \rightarrow\  0, \quad C=\mathrm{const.}
\label{eq:n6mrl}
\end{equation}
is the appropriate limit one has to perform in the non-linear integral
equations in order for the cross ratios to assume their limiting values
in the Regge regime. When the limit \eqref{eq:n6mrl} is taken, the
integrals in the NLIE \eqref{eq:ysys1}-\eqref{eq:ysys3} may be neglected.
Consequently, we obtain a set of explicit expressions for the form of the
$\Yf$-functions in the limiting regime:
$$
\Yf_1(\theta) \cong e^{-m \cosh(\theta -i \phi) - C} \ , \quad
\Yf_2(\theta) \cong e^{-m \sqrt{2} \cosh(\theta -i \phi)} \ , \quad
\Yf_3(\theta) \cong e^{-m \cosh(\theta -i \phi) + C} \ . \
$$
Recall that these expressions should only be used for $|\mathrm{Im} \theta|
< \pi/4$. Outside this fundamental strip one needs to apply the
recursion relations \eqref{eq:recursion} to bring the arguments
back into the strip. Before we insert these expressions into the
formulas \eqref{eq:spe_cr} let us  replace $m$ and $\phi$ by the
new variables
\begin{equation}
\epsilon=e^{-m\cos\phi},\quad w=e^{m\sin\phi},
\end{equation}
which behave as $\epsilon\rightarrow 0$ and $w\rightarrow
\mathrm{const.}$ in the limit \eqref{eq:n6mrl}. In terms
of these new parameters, the cross ratios \eqref{eq:spe_cr}
can be expanded as
\begin{flalign*}
&u_1=1-\left(w+\frac{1}{w}+2\cosh C\right)
\epsilon+\mathcal{O}(\epsilon^2),%
&&u_2=w\epsilon+\mathcal{O}(\epsilon^2),%
&&u_3=\frac{\epsilon}{w}+\mathcal{O}(\epsilon^2)\ .
\end{flalign*}
While $u_2$ and $u_3$ only involve the $\Yf$ function
$\Yf_2$ in the fundamental strip, we need to use the
recursion relation \eqref{eq:recursion} to find $u_1$.
Hence, all the three cross ratios indeed show their
Regge behavior \eqref{eq:mrlu2}, \eqref{eq:mrlu3}
and \eqref{eq:mrlu1}.

\subsection{Multi-Regge limit for $n=7$ gluons}

For the $7$-point amplitude, the cross ratios are given in
terms of the \Yf-functions as
\begin{align} \label{eq:un71}
u_{11}&=\U{2}{2},&\, u_{21}&=\U{2}{-2},&\, u_{31}&=\U{2}{0},\\[2mm]
u_{12}&=\U{1}{-3},&\, u_{22}&=\U{1}{-1},&\, u_{32}&=\U{1}{1}.
\label{eq:un72}
\end{align}
We are going to demonstrate that that the cross ratios obtained from
the $\Yf$-system for $n=7$ display multi-Regge behavior if we take
\begin{eqnarray}
 m_s \rightarrow \infty \quad  & , & \quad C_s = \mathrm{const.} \nn \\[2mm]
   \phi_1\rightarrow 0 \ \quad  & , & \quad \phi_2\rightarrow
   -\frac{\pi}{4}\ .  \label{eq:n7mrl}
\end{eqnarray}
Note that the limiting value for the second angle $\phi_2$ is
non-vanishing. In the subsequent section we shall argue that
in this limit, all integral contributions may be neglected and
that the prescription \eqref{eq:n7mrl} is the only one that
provides the correct Regge asymptotics of cross ratios. For
the moment let us just check how things work with the limit we
propose. As in the discussion for $n=6$ we shall switch
from variables $m_s$ and $\phi_s$ to
\begin{align}
\epsilon_1&=e^{-m_1\cos\phi_1},&\, w_1&=e^{m_1\sin\phi_1},\\[2mm]
\epsilon_2&=e^{-m_2\cos\left(\frac{\pi}{4}+\phi_2\right)},&\, w_2&=e^{m_2\sin\left(\frac{\pi}{4}+\phi_2\right)}.
\label{eq:epw2}
\end{align}
In the above mentioned limit \eqref{eq:n7mrl}, these quantities
behave as  $\epsilon_i\rightarrow 0$, $w_i\rightarrow\mathrm{const.}$
Let us begin our study of cross rations the simplest cases,
namely the two cross ratios
\begin{align}
u_{32}&=\U{1}{1}\cong\frac{e^{-\sqrt{2}m_1\cos
\left(\frac{\pi}{4}-\phi_1\right)}}{1+ e^{-\sqrt{2}m_1\cos
\left(\frac{\pi}{4}-\phi_1\right)}}
=\frac{\frac{\epsilon_1}{w_1}}{1+\frac{\epsilon_1}{w_1}}
=\frac{\epsilon_1}{w_1}+\mathcal{O}(\epsilon^2),\\[2mm]
u_{22}&=\U{1}{-1}\cong\frac{e^{-\sqrt{2}m_1\cos
\left(\frac{\pi}{4}+\phi_1\right)}}{1+ e^{-\sqrt{2}m_1
\cos\left(\frac{\pi}{4}+\phi_1\right)}}
=\frac{\epsilon_1w_1}{1+\epsilon_1w_1}=
\epsilon_1w_1+\mathcal{O}(\epsilon^2).
\end{align}
Up to this point, things work pretty much the same way as
for the cross ratios $u_2$ and $u_3$ in the case of $n=6$.
The next cross ratio we want to look at is
\begin{equation}
u_{31}=\U{2}{0}\cong\frac{e^{-\sqrt{2}m_2\cos\phi_2}}
{1+e^{-\sqrt{2}m_2\cos\phi_2}}=\frac{\frac{\epsilon_2}{w_2}}
{1+\frac{\epsilon_2}{w_2}}=\frac{\epsilon_2}{w_2}
+\mathcal{O}(\epsilon^2)\
\end{equation}
which is the first one to contain $\phi_2$. It is this last computation
that suggests for the first time to set the limiting value of $\phi_2$
to $\phi_2 = \pi/4$. If we had set $\phi_2 =0$, for example, we would
have been forced to omit the shifts by $\pi/4$ in the arguments of the
trigonometric functions in eq.\ \eqref{eq:epw2} in order to ensure
finiteness of $w_2$. Without the shifts, the Regge limit of $u_{31}$
would have been given by $\epsilon_2$. But since the definition
of $w_2$ and $\epsilon_2$ included a shift, we had to add and subtract
the limiting value $\pi/4$ of the angle $\phi_2$ in the argument
of the cosine. This is how we obtained the familiar looking Regge
asymptotics of $u_{31}$ even though the construction of the cross
ratio only involved $\Ynt{2}{2}{0}$ with a vanishing upper index.\\
All remaining cross ratios involve values of $\Yf$-functions outside
the fundamental strip so that we need to make repeated use of the
recursion relation. We see that
\begin{equation}
\nonumber\Ynt{2}{2}{-2}=
\frac{1+\Ynt{2}{1}{-1}}{\Ynt{2}{2}{0}
\left(1+\frac{1}{\Ynt{3}{2}{-1}}\right)
\left(1+\frac{1}{\Ynt{1}{2}{-1}}\right)}
\cong\frac{1+\epsilon_1w_1}{\frac{\epsilon_2}{w_2}
(1+\frac{e^{-C_2}}{\epsilon_2})(1+\frac{e^{C_2}}{\epsilon_2})}
=\epsilon_2w_2+\mathcal{O}(\epsilon^2)
\end{equation}
and therefore
\begin{align}
u_{21}&=\U{2}{-2}\cong\frac{\epsilon_2w_2}
{1+\epsilon_2w_2}
=\epsilon_2w_2+\mathcal{O}(\epsilon^2).
\end{align}
Analogously, we obtain
\begin{align}
\Ynt{2}{2}{2}\cong\frac{1+\frac{\epsilon_1}{w_1}}
{\frac{\epsilon_2}{w_2}\left(1+w_2e^{-C_2}\right)
\left(1+w_2e^{C_2}\right)},
\end{align}
from which we find that
\begin{align}
u_{11}=\U{2}{2}\cong 1-\left(w_2+\frac{1}{w_2}
+2\cosh C_2\right)\epsilon_2+\mathcal{O}(\epsilon^2).
\end{align}
The last cross ratio we need is given in terms of
$\Y{2}{1}{-3}$. The \Yf-functions appearing in the
recursion relation are given below (without the full calculation).
\begin{equation*}
\Ynt{3}{1}{-2}\cong\frac{1+e^{C_2}\epsilon_2}{e^{-C_1}
\epsilon_1\left(1+\frac{1}{\epsilon_1w_1}\right)},\quad%
\Ynt{1}{1}{-2}\cong\frac{1+e^{-C_2}\epsilon_2}{e^{C_1}
\epsilon_1\left(1+\frac{1}{\epsilon_1w_1}\right)}.
\end{equation*}
This then leads to
\begin{align*}
\Ynt{2}{1}{-3}&=\frac{1+\Ynt{2}{2}{-2}}{\Ynt{2}{1}{-1}
\left(1+\frac{1}{\Ynt{3}{1}{-2}}\right)\left(1+\frac{1}{\Ynt{1}{1}{-2}}\right)}
\cong\frac{ (1+\epsilon_2w_2)}{\epsilon_1 w_1\left(1+\frac{e^{-C_1}
\epsilon_1\left(1+\frac{1}{\epsilon_1w_1}\right)}{1+e^{C_2}\epsilon_2}\right)
\left(1+\frac{e^{C_1}\epsilon_1\left(1+\frac{1}{\epsilon_1w_1}\right)}
{1+e^{-C_2}\epsilon_2}\right)}
\end{align*}
and finally
\begin{equation}
u_{12}=\U{1}{-3}\cong 1-\left(w_1+\frac{1}{w_1}+2\cosh C_1\right)
\epsilon_1+\mathcal{O}(\epsilon^2).
\end{equation}
This shows that the choice of parameters indeed gives the right limits for the
cross ratios.

\subsection{Finding the correct limit}
\label{sec:limit}
Our analysis in the previous section was based on the claim that integral
contributions to the NLIE can be neglected in the limit \eqref{eq:n7mrl}.
Under this assumption we showed that our cross ratios display multi-Regge
behavior. On the other hand we also suggested that the limit
\eqref{eq:n7mrl} was uniquely fixed by these requirements. Let us now
discuss these claims in a bit more detail.\\
It is clear from the relations \eqref{eq:un71} and \eqref{eq:un72} that
for a cross ratio vanishing in the multi-Regge limit, the corresponding
value of the $\Yf$-function has to vanish, as well. Values of $\Yf$-functions
that appear in the cross ratios $u_{1\sigma}$, on the other hand, must
diverge in the multi-Regge limit. For example, from the limiting behavior
$u_{31}\rightarrow 0$ we conclude $\Ynt{2}{2}{0}
\rightarrow 0$. Therefore, $\mathrm{log}\Ynt{2}{2}{0}$ has to be large and
negative. The same follows for the value $\mathrm{log}\Ynt{2}{1}{-1}$ considering that
$u_{32}\rightarrow 0$ in the multi-Regge limit. Because of the way the
parameter $m_s$ enters into eq.\ (\ref{eq:ysys2}) we enforce the desired
behavior if we make $m_s$ very large. In this limit, the integrals
appearing in eqs.\ (\ref{eq:ysys1})-(\ref{eq:ysys3}) can be neglected
because, assuming real masses for the moment, their leading contribution
can be written schematically as
\begin{equation}
\int d\theta' K(\theta-\theta')\mathrm{log}\left(1+\Yf(\theta')\right)
\cong \int d\theta' K(\theta-\theta')\mathrm{log}
\left(1+e^{-m_s\cosh{\theta'}}\right)
\rightarrow 0,
\end{equation}
as $m_s\rightarrow\infty$. This remains true even after introducing
complex masses, as is shown in section \ref{sec:residues}. In this limit,
we can now study the behaviour of the \Yf-functions and find constraints
on the $\phi_s$ and $C_s$.\\
For further illustration we now wish to look at the cross ratios $u_{11}$
and analyze how we ensure $u_{11} \rightarrow 1$ in the multi-Regge limit.
The relevant value of the $\Yf$-function is given by
\begin{align}
\Y{2}{2}{2}\cong\frac{1+e^{-\sqrt{2}m_1\cos\left(\frac{\pi}{4}-\phi_1\right)}}
{e^{-\sqrt{2}m_2\cos\phi_2}\left(1+e^{m_2\cos\left(\frac{\pi}{4}-
\phi_2\right)-C_2}\right)\left(1+e^{m_2\cos\left(\frac{\pi}{4}-\phi_2\right)
+C_2}\right)}\rightarrow\infty.
\end{align}
If we want the $\Yf$-function to diverge, every term of the denominator
has to vanish in the multi-Regge limit. Writing the denominator as
\begin{equation}
e^{-\sqrt{2}m_2\cos\phi_2}+e^{\sqrt{2}m_2\sin\phi_2}+2\cosh C_2 e^{-m_2\cos\left(\frac{\pi}{4}+\phi_2\right)}
\end{equation}
we see that
\begin{equation}
\phi_2\in\left(-\frac{\pi}{2},\frac{\pi}{2}\right)
\cap\left(-\pi,0\right)\cap\left(-\frac{3\pi}{4},
\frac{\pi}{4}\right)=\left(-\frac{\pi}{2},0\right),
\end{equation}
at least if we assume that it will stay in the interval
between $-\pi \leq \phi_2 \leq \pi$. In order to find the
specific value that $\phi_2$ should assume, we recall from
eq.\ (\ref{eq:ratiou2u3}) that the ratio $u_{21}/u_{31}$
must remain constant in the multi-Regge limit. According
to our equations \eqref{eq:un71}, this requires
\begin{align}
\frac{\Ynt{2}{2}{-2}}{\Ynt{2}{2}{0}}\cong\frac{1}
{e^{-2\sqrt{2}m_2\cos\phi_2}
\left(1+e^{m_2\cos\left(\frac{\pi}{4}+\phi_2\right)
-C_2}\right)\left(1+e^{m_2\cos\left(\frac{\pi}{4}+
\phi_2\right)+C_2}\right)}\rightarrow\mathrm{const.}
\end{align}
Here we have also used that both $\Ynt{2}{2}{-2}$ and
$\Ynt{2}{2}{0}$ vanish in the multi-Regge limit. For
the ratio of these values to be constant, at least one
term of the denominator has to go to a constant with
the other terms in the denominator going to zero.
This suggests the two possible limits $\phi_2\rightarrow
-\frac{\pi}{4}$ or $\phi_2\rightarrow \frac{\pi}{2}$.
The latter limiting behavior, however, is excluded by
the constraint we derived from $u_{11}\rightarrow 1$.
A similar analysis is carried out for the remaining
$\Yf$-functions in appendix \ref{sec:7pt_values}. There
we derive all constraints one can put on the parameters
of the NLIE. These are solved by $\phi_1\rightarrow 0$
and $\phi_2\rightarrow -\frac{\pi}{4}$, as we anticipated
above. And indeed in section \ref{sec:ew} we saw that such
a limiting behavior of the phases $\phi_s$ produces the
correct multi-Regge behavior for all the cross ratios.\\
The most surprising outcome of our discussion for $n=7$ is
that the limiting values of the phases need no longer be zero,
in contrast to what we found for $n=6$. One may generalize the
arguments outlined here to larger number of gluons $n \geq 8$
to find that $\phi_s=-(s-1)\frac{\pi}{4}$ appear to be the
correct values for the phases in the multi-Regge limit. A
full proof of this fact, however, requires a closer look at
the $\Yf$-system for non-zero phases and possible residues.

\section{Multi-Regge limit of the TBA - general case}
\label{sec:npt}
As we have argued in the previous section, the phases $\phi_s$
may approach large values in the multi-Regge regime. Since the
original $\Yf$-system is only valid as long as phases satisfy
$|\phi_s-\phi_{s'}| < \pi/4$ we must be prepared to include
additional contributions that arise from poles in the kernel,
see our comments in section \ref{sec:ampy} and \cite{Alday:2010ku}. In
the following analysis we will make the ansatz
\begin{align}
&m_s\ \rightarrow \, \infty ,\\[2mm]
&\phi_s\ =\ -(s-1)\frac{\pi}{4},\label{eq:mrlphi}\\[2mm]
&C_s\ = \ \mathrm{const.}
\end{align}
for the multi-Regge limit of the parameters $m_s, \phi_s$ and
$C_s$. Our main goal is to show that in this limit, the cross
ratios \eqref{eq:cr1}-\eqref{eq:cr3} have multi-Regge behavior,
i.e.\ that they behave as described in eqs.\ \eqref{eq:mrlu2},
\eqref{eq:mrlu3} and \eqref{eq:mrlu1}. The contributions from
poles of the kernel functions turn out to play a vital role in
this analysis. Therefore, we begin with a few general comments
on the behavior of the \Yf-system for large masses in the
presence of large phases $\phi_s$. These then allow us to find
explicit values  for the physical cross ratios for an arbitrary
number of external gluons.

\subsection{Large phase residues and multi-Regge limit}
\label{sec:residues}
In this section we revisit the structure of the \Yf-system
equations in the presence of large phases $\phi_s$. Recall
that in the presence of complex masses the equations can
be written as
\begin{equation}
\mathrm{log}\tilde{\Yf}_{2,s}(\theta)=-\sqrt{2}|m_s|
\cosh\theta+\sum\limits_{a',s'}\int d\theta'
K^{2,a'}_{s,s'}(\theta-\theta'+i\phi_s-i\phi_{s'})
\mathrm{log}(1+\tilde{\Yf}_{a',s'}(\theta')),
\label{eq:ysysmod}
\end{equation}
where $\tilde{\Yf}_{a,s}(\theta)=\Yf_{a,s}(\theta+i\phi_s)$,
cf.\ \cite{Alday:2010vh}. We only displayed the equations for
$\tilde{\Yf}_{2,s}$ here, but similar equations obviously hold
for the other $\Yf$-functions, as well. Eqs.\ \eqref{eq:ysysmod}
determine the functions $\tilde \Yf$ in the fundamental strip
$ |\mathrm{Im} \theta| < \pi/4$ as long as the phase
differences satisfy $|\phi_s - \phi_{s'}| < \pi/4$ . For
$\mathrm{Im}(\theta+i\phi_s-i\phi_{s'})=k\cdot\frac{\pi}{4}$,
the kernels are singular and we have to pick up residues from
these poles\footnote{Note that we have to pick up a pole only if the kernel singularity is crossed. For values on the singularities, an $i0$-prescription can be used (cf.\cite{Alday:2010vh}).}. More precisely, the kernels $K_2$, $K_3$ become
singular for $k= 2n+1$, while $K_1$ is singular for $k=2(2n+1)$.
In other words, at least one of the kernel functions has a pole
whenever
$$ k \in {\mathcal X} := \{ 2\mathbb{Z} + 1\} \cup
\{2(2\mathbb{Z}+1)\} \ . $$
Once we include the residues of the pole contributions, the
integral equations are given by
\begin{align}
\nonumber
\mathrm{log} \tilde{\Yf}_{2,s}(\theta)&=
-\sqrt{2}|m_s|\cosh\theta + \sum_\nu n_\nu
\mathrm{log}\left(1+\tilde{\Yf}_{a_\nu,s_\nu}\left(\theta
+i\phi_s-i\phi_{s_\nu}-ik_\nu\frac{\pi}{4}\right)\right)\\[2mm]
& \quad \quad + \sum\limits_{a',s'}\int d\theta' K^{2,a'}_{s,s'}
(\theta-\theta'+i\phi_s-i\phi_{s'})\mathrm{log}
(1+\tilde{\Yf}_{a',s'}\left(\theta'\right))
\label{eq:ysys_fmod}
\end{align}
In eq.\ (\ref{eq:ysys_fmod}), it is understood that the sum in the
first line extends over all the relevant pole contributions. We will
discuss this in more detail below. As they stand, equations
\eqref{eq:ysys_fmod} are valid for arbitrary $\theta$ in the complex
plane. Of course, the number of terms in the summation over $\nu$
depends on the value of $\mathrm{Im}(\theta)$.\\
Before we proceed to analyze the behavior of eqs.\ \eqref{eq:ysys_fmod}
in the limit of large masses, let us have a closer look at the $\Yf$-system
for real argument $\theta$. Recall from section \ref{sec:ampy} that the kernels
$K^{a,a'}_{s,s'}$ are non-zero only for $s'=s$ or $s' = s\pm 1$. Hence,
the differences $\phi_s-\phi_{s'}$ that appear in our limit \eqref{eq:mrlphi}
are restricted to values $|\phi_s-\phi_{s'}| \leq \pi/4$. We conclude that,
as long as $\theta$ is real, no poles of the kernel functions are actually
crossed in the limit \eqref{eq:mrlphi} and hence the $\Yf$-functions
$\Yf_{a,s} (\theta), \theta \in \mathbb{R}$, obey the original $\Yf$-system
without additional residue terms.\\
This comment becomes important in passing to the large mass limit of
eqs.\ \eqref{eq:ysys_fmod}. Note that the integral in the second line
extends over $\theta'$ along the real line. Hence, to evaluate the
integral, we only need to know $\tilde \Yf_{a',s'}(\theta')$ for real
$\theta'$ where it is unaltered by pole contributions and consequently
still given by
$$ \tilde \Yf_{2,s'}(\theta') \cong e^{-\sqrt{2}|m_s|\cosh \theta'}
\quad ,\quad
\tilde \Yf_{a',s'}(\theta') \cong e^{- |m_s| \cosh \theta' + (a'-2)C_s}\quad\mathrm{for}\,a'=1,3,
$$
in the limit of large masses $|m_s|$ and for $\theta' \in \mathbb{R}$.
It follows once again that all integral contributions to eqs.\ \eqref{eq:ysys_fmod} can be neglected in the limit \eqref{eq:mrlphi}.
Shifting back to the \Yf-functions, we see that after neglecting the
integrals the equations look like\footnote{Note that the variable $\theta$ was shifted by
$-i\phi_s$ when we passed from $\tilde \Yf_{a,s}$ to $\Yf_{a,s}$.}
\begin{equation}
\mathrm{log} \Yf_{2,s}(\theta)= -\sqrt{2}|m_s|\cosh\left(\theta-i\phi_s\right)
+ \sum_\nu n_\nu \mathrm{log}\left(1+\Yf_{a_\nu,s_\nu}
\left(\theta-ik_\nu\frac{\pi}{4}\right)\right).
\label{eq:ysys_final}
\end{equation}
It is important now to describe the sum over pole contributions in
some more detail. To begin with all labels $a_\nu,s_\nu$ can only run
over nine possible values. While $a_\nu$ is free to assume any of its
three values, $s_\nu$ cannot deviate from $s$ by more than one unit, i.e.\
$$ a_\nu = 1,2,3 \quad , \quad s_\nu \in \{s,s\pm 1\}\ . $$
Given any such choice of $(a_\nu,s_\nu)$ we compute
the quantity $\kappa_{s_\nu} = \frac{4}{\pi}\mathrm{Im}(\theta - i\phi_{s_\nu})
$. It determines the possible values of the integer
$$
k_\nu \ \in \  (0, \kappa_{s_\nu}) \cap {\mathcal X} \ . $$
Given $(a_\nu,s_\nu)$ and $k_\nu$ we must still find the integer
$n_\nu$. It is determined by the form of the functions $\alpha_s,
\beta_s$  and $\gamma_s$ in eqs.\ \eqref{eq:alpha}-\eqref{eq:gammaY},
the coefficients of these functions in the $\Yf$-system
\eqref{eq:ysys1}-\eqref{eq:ysys3} and sign of $k_\nu$. We will
discuss some examples in the next subsection.

\subsection{Multi-Regge limit for $n=8$ gluons}

As we shall discuss at the end of the next subsection, pole contributions
from the kernels only start to enter the limit \eqref{eq:mrlphi} starting
from $n = 8$ external gluons. It is useful to analyze the $n=8$ case first
to illustrate the general formula \eqref{eq:ysys_final}. Generalizing the
variables we used in our analysis of the multi-Regge limit for 7 gluons
we introduce
\begin{align}
\epsilon_s &=e^{-m_s\cos\left((s-1)\frac{\pi}{4}+\phi_s\right)},\\
w_s &=e^{m_s\sin\left((s-1)\frac{\pi}{4}+\phi_s\right)},
\end{align}
where $s=1,2,3$. In the limit \eqref{eq:mrlphi} these variables
behave as $\epsilon_s\rightarrow 0$ and $w_s\rightarrow \mathrm{const}$.\\
As an example, let us analyze the limit of $\Ynt{2}{2}{-2}$. Since we want
to determine $\Yf_{2,2}$ at $\theta = - i \pi/2$ and $\phi_{s_\nu} = -
(s_\nu-1)\pi/4$, the parameter $\kappa_{s_\nu}$ defined at the end of the
previous subsection reads $\kappa_{s_\nu} = s_\nu-3$. In order for the
interval $(0,s_\nu-3)$ to have a non-vanishing intersection with
$\mathcal X$, we must have $s_\nu=1$. In the case at hand, the
intersection consists of a single point $k = -1$. Looking back at eq.\ \eqref{eq:ysys2} we note that only $K_2$ possesses a pole at $-i\pi/4$. The coefficient of $K_2$ in the equation for $\log \Yf_{2,2}$
is $-\alpha_2$. The latter contains only one contribution with $1=
s_\nu =s-1$, namely the term $-\log (1+ \Yf_{2,1})$. Consequently, we
find
\begin{equation}\label{eq:Y222res}
\Ynt{2}{2}{-2}= e^{-\sqrt{2} m_2 \cos(\pi/2+\phi_2)}
\cdot\left(1+\Ynt{2}{1}{-1}\right) = \epsilon_2 w_2 \cdot
\left(1+\Ynt{2}{1}{-1}\right).
\end{equation}
Note that in this particular case the correction term becomes trivial
in the limit \eqref{eq:mrlphi} since $\Ynt{2}{1}{-1} = \exp(-m_s)$.
Analogously, we can analyze the residue structure of the remaining
\Yf-functions. The result is shown in figure \ref{fig:8pt_res} which
we explain in the following.
\begin{figure}
\centering
\includegraphics{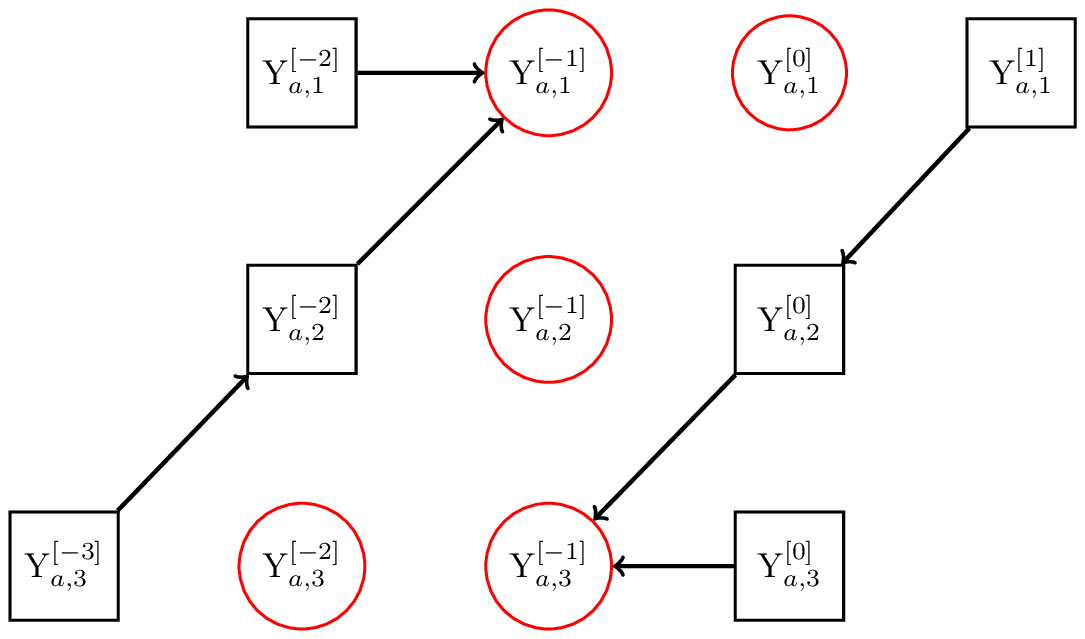}
\caption{Residue structure for \Yf-functions in 8-point case.}
\label{fig:8pt_res}
\end{figure}
Every node in the diagram represents the $3$ values $\Yf_{as}^{\left[k\right]}$
with $a=1,2,3$ while $s$ and $k$ are kept fixed. Encircled nodes correspond to
\Yf-functions that receive no corrections from residues. Arrows points to the
nodes from which a given \Yf-function receives residue terms. Our result
\eqref{eq:Y222res}, for example, is represented by the arrow that connects the
square box around $\Y{a}{2}{-2}$ with the circle around $\Y{a}{1}{-1}$.
Values of $\Y{a}{s}{r}$ that are not included in the figure not only receive
residues with $k_\nu = \pm 1$, but include higher values of $k_\nu$. Those
have a more complicated structure and are better bypassed through the use
of recursion relations, if possible. We see that with the help of the
\Yf-functions at $\theta=-i\pi/4$ we can determine all \Yf-functions at
$\theta=0$. Once we have constructed $\Y{a}{s}{-1}$ and $\Y{a}{s}{0}$,
all remaining values can be reconstructed with the help of the recursion
relation \eqref{eq:recursion}. A complete analysis gives the following
results for the cross ratios in the limit \eqref{eq:mrlphi},
\begin{align}
&u_{1\sigma}=1-\left(w_{n-4-\sigma}+\frac{1}{w_{n-4-\sigma}}+2\cosh C_{n-4-\sigma}\right)\epsilon_{n-4-\sigma},\\[2mm]
&u_{2\sigma}\ =\ \epsilon_{n-4-\sigma}w_{n-4-\sigma},\\[2mm]
&u_{3\sigma}\ = \ \epsilon_{n-4-\sigma}w_{n-4-\sigma}^{-1}.
\end{align}
So, we find that all cross ratios have multi-Regge asymptotics
\eqref{eq:mrlu2}, \eqref{eq:mrlu3} and \eqref{eq:mrlu1}.

\subsection{The general case of n-gluon scattering}
\label{sec:gen_case}
After discussing 7- and 8-gluon amplitudes we finally proceed to the
general case. Our goal is to show that the cross ratios \eqref{eq:cr1}-\eqref{eq:cr3} that are obtained from the $\Yf$-functions through
eqs.\ \eqref{eq:spe_cr} show multi-Regge behaviour \eqref{eq:mrlu2},
\eqref{eq:mrlu3} and \eqref{eq:mrlu1} in the limit \eqref{eq:mrlphi}.
As before, we can study the residue structure of the \Yf-system for
phases $\phi_s = - (s-1)\pi/4$. The results are encoded in
figure \ref{fig:npt_res}.
\begin{figure}
\centering
\includegraphics[scale=.95]{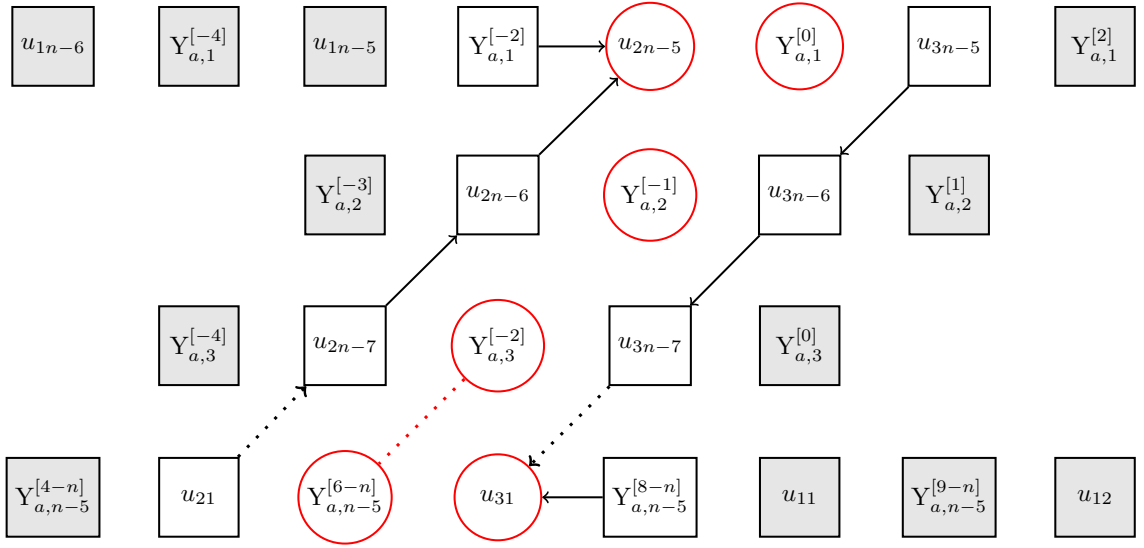}
\caption{Residue structure for n points. \Yf-functions not shown in the figure correspond to gray boxes.}
\label{fig:npt_res}
\end{figure}
Gray nodes correspond to \Yf-functions with a more complicated residue
structure. We explain in appendix \ref{sec:res_grey} how our pattern can
be used to read off the resulting contributions. If a cross ratio
\eqref{eq:spe_cr} is obtained from the \Yf-function $\Y{2}{s}{k}$, we
have indicated this by putting the cross ratio into the node instead
of the \Yf-function. We have already used the shift symmetry to put
some of the $u_{1\sigma}$ in the first row.\\
We see that the cross ratios $u_{2\sigma}$ and $u_{3\sigma}$ lie on a
diagonal whose `\textit{residue flow}' stays within the diagonal. The
endpoints of the residue flow are given by $u_{31}$ and $u_{2\, n-5}$,
which are free of residues and can be calculated directly to to give
\begin{align}
u_{31}&=\epsilon_{n-5}w_{n-5}^{-1},\\[1mm]
u_{2\, n-5}&=\epsilon_1 w_1.
\end{align}
It is not difficult to determine the remaining cross ratios
$u_{2\sigma}$ and $u_{3\sigma}$. Recall from eq.\ \eqref{eq:spe_cr}
that the cross ratios $u_{2\sigma}$ are given by
\begin{equation}
u_{2\sigma}\ = \ \U{n-4-\sigma}{-(n-4-\sigma)}.
\end{equation}
According to our general expression \eqref{eq:ysys_final}, the
values of the $\Yf$-functions that enter these cross ratios are
given by
\begin{equation}
\Ynt{2}{s}{-s}=e^{-\sqrt{2}m_s\cos\left(\frac{\pi}{4}+(s-1)
\frac{\pi}{4}+\phi_s\right)}\cdot\left(\mathrm{Residue}\,
\mathrm{terms}\right)=\epsilon_s w_s\cdot
\left(\mathrm{Residue}\,\mathrm{terms}\right).
\end{equation}
We can conclude by iteration that the residue terms along this
diagonal only introduce corrections of the order $\epsilon^2$
or higher and can be neglected. Hence, we obtain
\begin{equation}
u_{2\sigma}=\epsilon_{n-4-\sigma}w_{n-4-\sigma}.
\end{equation}
One can go through the same arguments to find
\begin{equation}
u_{3\sigma}= \epsilon_{n-4-\sigma}w_{n-4-\sigma}^{-1}\ .
\end{equation}
This finally leaves us with the cross ratios $u_{1\sigma}$ for
which we have to work a little bit harder, as these variables all
have a more complicated residue structure. We will make use of an
inductive argument. Going from a $n$-point amplitude to a $(n+1)$-point
amplitude introduces another row in our above pattern. However, the
old \Yf-functions do not change their values as long as their
residues are not affected by the new row. What changes, however,
is the location of the cross ratios in the pattern. The most
important change for our purposes is that all $u_{1\sigma}$ in
the first row are shifted by two boxes to the left in our pattern,
and a new cross ratio $u_{1((n+1)-5)}$ appears which is related
to $\Y{2}{1}{-3}$. This means, for example, that $u_{1((n+1)-5)}$
in the $(n+1)$-point amplitude will take the value of $u_{1n-5}$ in
the $n$-point amplitude, since all residue are unaffected by the
new row. Following our discussion in appendix \ref{sec:res_grey},
it is clear that a \Yf-function will keep its $n$-point value if
the diagonal from its node to the lower right hits the $(n+1)$-point
$u_{2\sigma}$-diagonal at the position of $u_{22}$, or above.
It turns out that the last element, i.e. the element with smallest
value of $\theta$, in the first row that keeps its value is given
by $u_{13}$. This means, that all $u_{1\sigma}$ for $\sigma=3,...,
n-4$ can be calculated using the known $n$-point values. However,
$u_{11}$ and $u_{12}$ are then given by the symmetry (\ref{eq:symn}) relating
$u_{1\sigma}\leftrightarrow u_{1(n-4-\sigma)}$. We have established
above that
\begin{equation}
u_{1\sigma}=1-\left(w_{n-4-\sigma}+\frac{1}{w_{n-4-\sigma}}+2\cosh C_{n-4-\sigma}\right)\epsilon_{n-4-\sigma}
\end{equation}
for the $7$- and $8$-point amplitude. By the previous argument, the
same remains true for all values of $n$. This shows that the $n$-point
solution is a $(n-5)$-fold copy of the $6$ point solution. In conclusion,
we now confirmed our identification of the multi-Regge regime with the
limit \eqref{eq:mrlphi} of the parameters in the $\Yf$-system.

\subsection{The 7-point case revisited}
\label{sec:7_case}
In the light of the general result derived in the last section, let
us revisit the 7-point case to study explicitly the equations governing
the Y-system in the fundamental strip and to understand why no large
residues had to be considered in our earlier analysis. Recall that for
7-points the multi-Regge limit corresponds to taking $m_s$ large, $C_s$
constant, $\phi_1=0$ and $\phi_2=- \pi/4$. For $-\pi/4<\mathrm{Im}
\theta\leq 0$ we have that $\mathrm{Im}\left(\theta-i\phi_s\right)\in
\left(-\pi/4,\pi/4\right)$. Since none of these terms crosses a multiple
of $\pi/4$, no singularities arise and we do not have to pick up any residues.
Therefore, the Y-system (\ref{eq:ysysmod}) is valid without any modifications.
For $0<\mathrm{Im}\theta< \pi/4$, however, $\mathrm{Im}\left(\theta-i\phi_2
\right)$ crosses $\pi/4$ and we have to pick up residues. Following the
procedure outlined in the previous subsection, we find the following
equations:
\begin{align*}
\mathrm{log}\Yf_{11}\left(\theta\right)=&-|m_1|\cosh\left(\theta-i\phi_1\right)+\sum\limits_{a',s'}\int d\theta' K_{1,s'}^{1,a'}\left(\theta-\theta'-i\phi_{s'}\right)\mathrm{log}\left(1+\Yf_{a',s'}\left(\theta'+i\phi_{s'}\right)\right)\\[1mm]
&- C_1 + \mathrm{log}\left(1+\Yf_{12}\left(\theta-i\frac{\pi}{4}\right)\right),\\[2mm]
\mathrm{log}\Yf_{21}\left(\theta\right)=&-\sqrt{2}|m_1|\cosh\left(\theta-i\phi_1\right)+\sum\limits_{a',s'}\int d\theta' K_{1,s'}^{2,a'}\left(\theta-\theta'-i\phi_{s'}\right)\mathrm{log}\left(1+\Yf_{a',s'}\left(\theta'+i\phi_{s'}\right)\right)\\[1mm]
&+\mathrm{log}\left(1+\Yf_{22}\left(\theta-i\frac{\pi}{4}\right)\right),\\[2mm]
\mathrm{log}\Yf_{31}\left(\theta\right)=&-|m_1|\cosh\left(\theta-i\phi_1\right)+\sum\limits_{a',s'}\int d\theta' K_{1,s'}^{3,a'}\left(\theta-\theta'-i\phi_{s'}\right)\mathrm{log}\left(1+\Yf_{a',s'}\left(\theta'+i\phi_{s'}\right)\right)\\[1mm]
&+C_1 +\mathrm{log}\left(1+\Yf_{32}\left(\theta-i\frac{\pi}{4}\right)\right),\\[2mm]
\mathrm{log}\Yf_{12}\left(\theta\right)=&-|m_2|\cosh\left(\theta-i\phi_2\right)+\sum\limits_{a',s'}\int d\theta' K_{2,s'}^{1,a'}\left(\theta-\theta'-i\phi_{s'}\right)\mathrm{log}\left(1+\Yf_{a',s'}\left(\theta'+i\phi_{s'}\right)\right)\\[1mm]
&-C_2 -\mathrm{log}\left(1+\Yf_{22}\left(\theta-i\frac{\pi}{4}\right)\right),\\[2mm]
\mathrm{log}\Yf_{22}\left(\theta\right)=&-\sqrt{2}|m_2|\cosh\left(\theta-i\phi_2\right)+\sum\limits_{a',s'}\int d\theta' K_{2,s'}^{2,a'}\left(\theta-\theta'-i\phi_{s'}\right)\mathrm{log}\left(1+\Yf_{a',s'}\left(\theta'+i\phi_{s'}\right)\right)\\[1mm]
&-\mathrm{log}\left(1+\Yf_{12}\left(\theta-i\frac{\pi}{4}\right)\right)-\log\left(1+\Yf_{32}\left(\theta-i\frac{\pi}{4}\right)\right),\\[2mm]
\mathrm{log}\Yf_{32}\left(\theta\right)=&-|m_2|\cosh\left(\theta-i\phi_2\right)+\sum\limits_{a',s'}\int d\theta' K_{2,s'}^{3,a'}\left(\theta-\theta'-i\phi_{s'}\right)\mathrm{log}\left(1+\Yf_{a',s'}\left(\theta'+i\phi_{s'}\right)\right)\\[2mm]
&+C_2 -\mathrm{log}\left(1+\Yf_{22}\left(\theta-i\frac{\pi}{4}\right)\right).
\end{align*}
Let us now examine the $\Yf$-functions appearing in the residue terms.
To be specific, we will focus on the term involving $\Yf_{12}\left(\theta
-i\pi/4\right)$ in the first equation. For the argument of this $\Yf$-function
we have $-\pi/4<\mathrm{Im}\left(\theta-i \pi/4\right)<0$. As argued before, in
this range of the argument no residue terms appear and in the multi-Regge limit
the $\Yf$-function is given by
\begin{equation}
\Yf_{12}\left(\theta-i\frac{\pi}{4}\right)\cong e^{-|m_2|\cosh\left(\theta-i\frac{\pi}{4}-i\phi_2\right)-C_2}
=e^{-C_2}\epsilon_2^{\cosh\theta}w_2^{i\sinh\theta}.
\end{equation}
The exponent of $\epsilon_2$ is positive in the chosen range of $\theta$.
Therefore, the function goes to zero in the multi-Regge limit and the residue
containing $\Yf_{12}$ gives a negligible contribution. Analogously, we can
analyze the remaining residue terms. It turns out that they are all
negligible. The same is actually true for $n=8$ external gluons. However,
starting from 9-points the residue terms give significant contributions
in the fundamental strip, see Appendix \ref{sec:9pt_bethe} for explicit expressions.

\section{Bethe ansatz}
\label{sec:ba}

The analysis outlined above has shown that the non-linear integral equations
that control the n-gluon amplitudes at strong coupling simplify drastically
when we take the multi-Regge limit. In the limiting regime we can actually
neglect the integral contributions, possibly after taking some residue terms
into account. Such a limit is well known in the theory of integrable systems.
It corresponds to a large volume limit in which the solution of the integrable
model boils down to solving a set of algebraic Bethe ansatz equations.\\
Before we recall how the Bethe ansatz equations emerge in the multi-Regge
limit, we need to comment a bit more on the residue terms we picked up while
sending the phases $\phi_s$ to their limiting values $\psi_s = (1-s)\pi/4$.
The appearance of such residue terms has been discussed for the $AdS_3$
$\Yf$-system in Appendix B of \cite{Alday:2010vh}. Using experience from
closely related wall-crossing phenomena (see \cite{Gaiotto:2009hg}), the
authors of \cite{Alday:2010vh} demonstrated how residue terms can be
absorbed in a redefinition of the $\Yf$-functions. While bringing the
equations back into the standard form of a $\Yf$-system,
\begin{equation}
- \log \tilde \Yf_{A} (\theta) =  p_{A}(\theta) + \sum_B
   \int \dfrac{d\theta'}{2\pi } K_{AB}(\theta-\theta')
   \log \left(1 + \tilde \Yf_{B}(\theta')\right)\ \ ,
\label{TBA}
\end{equation}
it is necessary to introduce additional $\Yf$-functions. This implies
that the new set of equations \eqref{TBA} might involve more than the
$3n-15$ $\Yf$-functions we started with. The complete set of $\Yf$-functions
is enumerated by the label $A,B, \dots$. Constructing the source terms
$p_{A}$ and the kernel functions $K_{AB}$ is part of the task one has
to address while passing from a modified $\Yf$-system with residue terms
back to a  $\Yf$-system of the form \eqref{TBA}. We defer a detailed
 discussion of this procedure for the $AdS_5$ $\Yf$-system at $\phi_s
 = (1-s) \pi/4$ to a future publication. Let us only mention that the
 source terms $p_A$ for the new $\Yf$-functions that are introduced
 while rewriting the equations
possess the canonical form with masses $m_A$ that can be obtained from
the $n-5$ masses $m_s$ of the original $\Yf$-system.\\
Once we accept that the original $\Yf$-system with large phases
$\phi_s = (1-s)\pi/4$ can be brought into the form \eqref{TBA}, we
are prepared to review how Bethe ansatz equations emerge. In order
to do so, we represent the kernel functions $K_{AB}$ through new
objects $S_{AB}$,
\begin{equation}
 K_{AB}(\theta) = - \partial_\theta \log S_{AB}(\theta) \ \ .
\label{KfromS}  \end{equation}
As observed first by Dorey and Tateo \cite{Dorey:1996re,Dorey:1997rb},
upon analytic continuation of the parameters some of the solutions
to the equations $\tilde \Yf_{A}(\theta) = -1$ may cross the real axis.
We shall enumerate those solutions by an index
$\nu = 1, \dots, N_{A}$:
\begin{equation} \label{poleeq}
 \tilde \Yf_{A}(\theta^{(A)}_\nu) = -1 , \ \ \mbox{ for } \ \
 \nu = 1,\dots, N_{A}
  \ \ .
\end{equation}
When this happens, the integral on the right-hand side of equation
\eqref{TBA} picks up a residue term since there is a pole crossing
the integration contour. Hence, after analytic continuation the
equations \eqref{TBA} take the form
\begin{eqnarray}
- \log \tilde \Yf_{A}(\theta) & = &
p_{A}(\theta) +
\sum_{B} \sum_{\nu = 1}^{N_{B}}
\log S_{AB}(\theta-\theta^{(B)}_\nu)  \nonumber \\
 & & +
\sum_{B} \int \dfrac{d\theta'}{2\pi }
   K_{AB}(\theta-\theta')
   \log \left(1 + \tilde \Yf_{B}
   (\theta')\right)\ \ .
\label{TBAex}
\end{eqnarray}
If we now send the parameters of our nonlinear integral equations
back into a regime where the integrals can be neglected, e.g. into
the multi-Regge regime we explored in this note, the equations
\eqref{TBAex} become
\begin{equation}
- \log \tilde \Yf_{A}(\theta) = p_{A}(\theta) +
\sum_{B} \sum_{\nu = 1}^{N_{B}}
\log S_{AB}(\theta-\theta^{(B)}_\nu) \ \ .
\label{TBAlogBA}
\end{equation}
We can exponentiate this set of equations for the functions $\Yf_{A}(\theta)$
and insert the values $\theta = \theta^{(A)}_\nu$ satisfying eq.\ \eqref{poleeq}
to obtain
\begin{equation}
e^{ i R k_{A}(\theta^{(A)}_\mu)}
= \prod_{B} \prod_{\nu = 1}^{N_{B}}
S_{AB}(\theta^{(A)}_\mu -
\theta^{(B)}_\nu) \ \
\label{TBABA}
\end{equation}
with $k_A(\theta) := i p_A(\theta)/R$. The parameter $R$ and the
functions $k_A$ were introduced here to help interpreting the
equations \eqref{TBABA}. In our context, these equations simply
determine the possible location of the solutions $\theta^{(A)}_\nu$
to the equations \eqref{poleeq}. But the form of the equations
coincides with the usual Bethe ansatz that imposes single-valuedness
of wave functions for particles on a 1-dimensional circle of
circumference $R$. The term $\exp(iR k_{A})$ accounts for the phase
shift of a freely moving particle with momentum $ k_{A}(\theta^{(A)}_\nu)$
when we take it once around the circle. The remaining factors arise
from the scattering with other particles that may be distributed
along the 1-dimensional circle. Hence, the quantities $S_{AB}$
introduced in eq.\ \eqref{KfromS} are interpreted as a scattering
matrix for excitations of some integrable system and the source
terms $k_{A}$ describe the momentum.\\
To make the Bethe ansatz equations \eqref{TBABA} for the multi-Regge
limit of the bubble ansatz more explicit, we need to determine the
range of the label $A$, the source terms $p_A(\theta)$ and the
kernel functions $K_{AB}(\theta)$, or rather the corresponding
S-matrices. Only in the case of $n=6$ external gluons these can
be read off easily from the original $\Yf$-system. The general case
will be addressed in a future publication. Much of the above would
have remained valid if we had not brought to the modified $\Yf$-system
back into the form \eqref{TBA}. But the resulting Bethe ansatz equations
\eqref{TBABA} would have been modified as well, with the left hand side
being replaced by a sum of products of `phases' $\exp(iR k_{a,s})$.
Examples can be worked out from our formulas for the multi-Regge limit
of the $\Yf$-system with $n=9$, see appendix \ref{sec:9pt_bethe}. Such modified Bethe
ansatz equations can certainly be studied numerically. Nevertheless,
we believe that a deeper understanding of the underlying integrable
system requires to absorb the residue terms so that the Bethe ansatz
equations take the standard form.
\\
Let us finally discuss the form of the free energy \eqref{eq:Afree}.
As explained in \cite{Alday:2010vh} the original expression remains
valid in the presence of large phases $\phi_s$, but it needs to be
rewritten in terms of the $\Yf$-functions $\tilde \Yf_A$. Upon
analytic continuation of the parameters, the \Yf-functions can give
rise to pole terms that cross the real axis. This happens precisely
when the conditions \eqref{poleeq} are satisfied. After taking the
multi-Regge limit, only these pole terms survive and one should
obtain an expression of the form
\begin{equation}
A_\mathrm{free}  =   \sum_{A} \sum_{\nu=1}^{N_{A}}
e_A( \theta^{(A)}_\nu)  \ .
\label{eq:AfreeMRL}
\end{equation}
with some energy functions $e_A= e_A(\theta)$ that must be determined
while passing from the modified $\Yf$-system with residue terms to eqs.\
\eqref{TBA}. Hence, in order to evaluate scattering amplitudes of
strongly coupled \N4 SYM theory in the multi-Regge limit, we have to
solve the Bethe ansatz equations \eqref{TBABA} for the positions
$\theta^{(A)}_{\nu}$ of the solutions to eq.\ (\ref{poleeq}). Once
these are found, we can easily evaluate $A_\mathrm{free}$.

\section{Conclusions and outlook}

In this note we studied the multi-Regge limit of scattering amplitudes
in strongly coupled \N4 SYM theory for any number $n$ of
external gluons. As reviewed above, the remainder function $R^{(n)}$
is determined through an auxiliary 1D quantum system \eqref{eq:ysys1}-%
\eqref{eq:ysys3} which depends on $3n-15$ parameters $m_s, \phi_s$ and
$C_s$. The latter map in a complicated way to the $3n-15$ independent
cross ratios that parametrize the scattering process. Our
central result \eqref{eq:mrlphi} identifies the values of the parameters
in the 1D quantum system that correspond to the multi-Regge limit of the
4-dimensional gauge theory. Since the relevant limit \eqref{eq:mrlphi}
involves sending all the mass parameters $m_s$ to infinity, the
multi-Regge (high-energy) regime of the gauge theory maps to the large
volume (low energy) limit of the 1D quantum system. In such a limit, the
1D quantum system simplifies drastically. More precisely, the non-linear
integral equations \eqref{eq:ysys1}-\eqref{eq:ysys3} can be replaced by
a much simpler set of algebraic equations \eqref{TBABA}.\\
We find it remarkable that the computation of scattering amplitudes
simplifies at both weak and strong coupling. In the quest for the exact
S-matrix of \N4 SYM theory, one that interpolates all the way from weak
to strong coupling, it could therefore pay off to consider the
multi-Regge regime as an intermediate step before addressing general
kinematics. Our results suggests that the multi-Regge regime could be
tractable even for finite coupling, at least more tractable than the
full dependence of the remainder functions on all the $3n-15$ cross
ratios. On the other hand, the Regge-limit imposes very strong
constrains on the analytical structure of the remainder functions
$R^{(n)}$. Explicit formulas for the Regge-limit of the remainder
functions could therefore be an important ingredient in reconstructing
the full scattering amplitude from more basic data.\\
But before thinking about the interpolation to finite coupling, there
are a few more immediate issues to be addressed. One is related to
wall-crossing phenomena we briefly mentioned in section \ref{sec:ba}. Recall
that our limit \eqref{eq:mrlphi} involves large phases $\phi_s$ in
which differences $|\phi_s - \phi_{s\pm1}|$ assume the critical value
$\pi/4$. As we described in much detail, these large phases bring
additional residue terms into the non-linear integral equations of
the $\Yf$-system. This happens starting from $n=7$ external gluons.
The corresponding modified equations can be obtained through an
algorithm we outlined in section \ref{sec:gen_case}. The simplest non-trivial
example was spelled out explicitly in subsection \ref{sec:7_case}. Following
a procedure that is inspired by the study of wall-crossing phenomena
\cite{Gaiotto:2009hg,Gaiotto:2012rg}, it should be possible to bring the
modified $\Yf$-system into the standard form \eqref{TBA}. Our discussion
in section \ref{sec:ba} assumed that the necessary steps have been carried out
already. But in order to determine the precise range of the labels $A,B$,
the momenta $p_A(\theta)$ and the S-matrix elements $S_{AB}$ that
enter eq.\ \eqref{TBABA} for $n>6$ external gluons, one cannot
avoid a detailed analysis. We leave this to future research.\\
Finally, we need to establish a map between the analytic continuation of the
kinematic variables and the numbers $N^A$ of Bethe roots $\theta^{(A)}_\nu$
in the previous section. In the case of the hexagon, such a correspondence
was determined through numerical studies. The authors of \cite{Bartels:2010ej}
found that the analytical continuation from the so-called physical to the
mixed regime (see \cite{Bartels:2010ej} for precise definitions) makes
two solutions of eq.\ \eqref{poleeq} cross the real axis. Hence, the
multi-Regge limit of the $n=6$ amplitude in the mixed regime corresponds
to the energy of a doubly excited state in the 1D quantum system at
infinite volume $R$. This agrees nicely with the analysis in the weakly
coupled theory. For the Regge-limit of the full 2-loop hexagon remainder
function $R^{(6)}$ to be non-zero, one needs to pass into the same mixed
regime that is associated with a non-trivial doubly excited state of the
Bethe ansatz equation \eqref{TBABA}. It is clearly desirable to extend
such studies beyond the case of the hexagon. Note that for larger number
of external gluons there exist many different regimes with non-vanishing
Regge-limit which probe eigenvalues of the BKP Hamiltonian \cite{BKP} with increasing
number of sites \cite{Bartels:2011nz}. We expect that such different regimes are associated with solutions of eqs. \eqref{TBABA} with an increasing number $N^A$ of Bethe
roots.
\bigskip

\noindent
{\bf Acknowledgments:} We wish to thank Patrick Dorey, Nicolay Gromov, Paul Heslop, Andrey Kormilitzin, Jan Kotanski, Lev Lipatov, J\"org Teschner, Pedro Vieira and Gang Yang for valuable discussions. This work was supported in part by the SFB 676.

\begin{appendix}

\section{Explicit values of \Yf-functions for 7-point amplitude}
\label{sec:7pt_values}

This appendix contains the complete analysis of the restrictions
that the desired multi-Regge behavior of cross ratios imposes on
the limiting values of the angles $\phi_s$ for $n=7$ gluons. Some
part of the required analysis was included and explained in
section \ref{sec:limit}. Here we simply state the remaining set of formulas
without further comments. Let us begin with the simplest cross
ratios whose evaluation does not require any use of the
recursion relation \eqref{eq:recursion}:
\begin{flalign*}
&u_{31}\rightarrow 0\implies \Ynt{2}{2}{0}\cong
e^{-\sqrt{2}m_2\cos\phi_2}\rightarrow 0 &&\Rightarrow \boxed{\phi_2\in\left(-\frac{\pi}{2},\frac{\pi}{2}\right)}&&\\[2mm]
&u_{32}\rightarrow 0\implies \Ynt{2}{1}{1}\cong e^{-\sqrt{2}m_1\cos\left(\frac{\pi}{4}-\phi_1\right)}\rightarrow 0 &&\Rightarrow \boxed{\phi_1\in\left(-\frac{\pi}{4},\frac{3\pi}{4}\right)}&&\\[2mm]
&u_{22}\rightarrow 0\implies \Ynt{2}{1}{-1}\cong e^{-\sqrt{2}m_1\cos\left(\frac{\pi}{4}+\phi_1\right)}\rightarrow 0 &&\Rightarrow \boxed{\phi_1\in\left(-\frac{3\pi}{4},\frac{\pi}{4}\right)}&&
\end{flalign*}
Since the remaining cross ratios involve values of the $\Yf$-functions outside
the fundamental strip we must use the recursion relations \eqref{eq:recursion}.
This gives:
\begin{flalign*}
\nonumber &u_{21}\rightarrow 0 &&\implies &&\Ynt{2}{2}{-2}&&=
\frac{1+\Ynt{2}{1}{-1}}{\Ynt{2}{2}{0}
\left(1+\frac{1}{\Ynt{3}{2}{-1}}\right)\left(1+\frac{1}{\Ynt{1}{2}{-1}}\right)}&&\\
&\quad&&\quad&&\quad&&\cong\frac{e^{\sqrt{2}m_2\cos\phi_2}}
{\left(1+e^{m_2\cos\left(\frac{\pi}{4}+\phi_2\right)-C_2}\right)
\left(1+e^{m_2\cos\left(\frac{\pi}{4}+\phi_2\right)+C_2}\right)}
\rightarrow 0&&\\[2mm]
&\quad&&\quad&&\quad&&\Rightarrow \boxed{\phi_2\in\left(
-\pi,0\right)\cup\left(\frac{\pi}{2},\pi\right)}&&\\[2mm]
&u_{11}\rightarrow 1&&\implies &&\Ynt{2}{2}{2}&&=\frac{1+\Ynt{2}{1}{1}}{\Ynt{2}{2}{0}
\left(1+\frac{1}{\Ynt{3}{2}{1}}\right)\left(1+\frac{1}{\Ynt{1}{2}{1}}\right)}&&\\
&\quad&&\quad&&\quad&&\cong\frac{e^{\sqrt{2}m_2\cos\phi_2}}
{\left(1+e^{m_2\cos\left(\frac{\pi}{4}-\phi_2\right)-C_2}\right)
\left(1+e^{m_2\cos\left(\frac{\pi}{4}-\phi_2\right)+C_2}\right)}
\rightarrow\infty&&\\[2mm]
&\quad&&\quad&&\quad&&\Rightarrow \boxed{\phi_2\in\left(-\frac{\pi}{2},0\right)}&&
\end{flalign*}
To determine the precise values of $\phi_s$ in the multi-Regge limit,
we look at the ratios \eqref{eq:ratiou2u3}
\begin{flalign*}
&\frac{u_{22}}{u_{32}}\rightarrow \mathrm{const.}&&\implies &&\frac{\Ynt{2}{1}{-1}}{\Ynt{2}{1}{1}}&&=
\frac{e^{-\sqrt{2}m_1\cos\left(\frac{\pi}{4}+\phi_1\right)}}
{e^{-\sqrt{2}m_1\cos\left(\frac{\pi}{4}-\phi_1\right)}}=e^{2m_1\sin\phi_1}\rightarrow \mathrm{const.}\Rightarrow \boxed{\phi_1\rightarrow 0}&&\\
&\frac{u_{21}}{u_{31}}\rightarrow\mathrm{const.}&&\implies &&\frac{\Ynt{2}{2}{-2}}{\Ynt{2}{2}{0}}&&=\frac{e^{2\sqrt{2}m_2\cos{\phi_2}}}{
\left(1+e^{m_2\cos\left(\frac{\pi}{4}+\phi_2\right)-C_2}\right)
\left(1+e^{m_2\cos\left(\frac{\pi}{4}+\phi_2\right)+C_2}\right)}
\rightarrow\mathrm{const.}&&\\[2mm]
&\quad&&\quad&&\quad&&\Rightarrow \boxed{\phi_2\rightarrow -\frac{\pi}{4}}&&
\end{flalign*}

\section{Residue structure for arbitrary Y-functions}
\label{sec:res_grey}
As mentioned in the text, the residue structure for Y-functions with large shifts is intricate.
In this section, we will demonstrate how the residue structure can be read off from the pattern introduced in the text. To do so, we will look at the specific example $\Ynt{2}{1}{-4}$.
The phases that appear in the Y-system equation are $\phi_1=0$ and $\phi_2=- \pi/4$.
Since we need to evaluate our $\Yf$-function at $\theta = - i\pi$, the quantity $\theta-i\phi_1$ crosses the poles $-i\frac{\pi}{4},-i\frac{\pi}{2},-i\frac{3}{4}\pi$, while $\theta-i\phi_2$ crosses $-i\frac{\pi}{4},-i\frac{\pi}{2}$. The resulting residue structure therefore reads
\begin{equation}
\Ynt{2}{1}{-4}=\epsilon_1^{-\sqrt{2}}\cdot\underbrace{\frac{\left(1+\Ynt{2}{2}{-3}\right)}
{\left(1+\Ynt{3}{1}{-3}\right)\left(1+\Ynt{1}{1}{-3}\right)}}_{\mathrm{Residues}\, \mathrm{at}\,-i\frac{\pi}{4}}\cdot\underbrace{\frac{\left(1+\Ynt{3}{2}{-2}\right)
\left(1+\Ynt{1}{2}{-2}\right)}{\left(1+\Ynt{2}{1}{-2}\right)^2}}_{\mathrm{Residues}
\,\mathrm{at}\,-i\frac{\pi}{2}}\cdot\underbrace{\frac{1}{\left(1+\Ynt{3}{1}{-1}\right)
\left(1+\Ynt{1}{1}{-1}\right)}}_{\mathrm{Residues}\,\mathrm{at}\,-i\frac{3}{4}\pi}.
\end{equation}
Graphically, this residue structure can be represented as in figure \ref{fig:res_large}, which remains valid for $\Ynt{a}{1}{-4}$.
\begin{figure}
\centering
\includegraphics{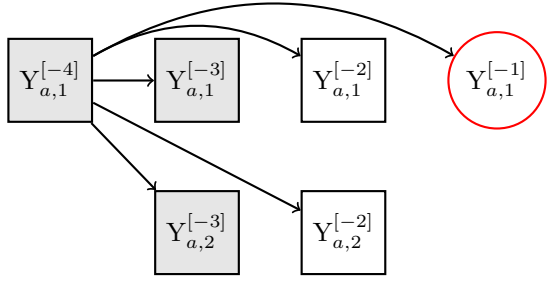}
\caption{Residue structure of $\Ynt{a}{1}{-4}$. For simplicity,
the residue dependencies of the other Y-functions are not shown.}
\label{fig:res_large}
\end{figure}
Of course, some of the \Yf-functions that appear as residues
themselves receive corrections from residue terms, for example
$\Ynt{2}{2}{-3}$ in the above example. However, it should be
clear from the graphical representation that the residue term
with the highest $s'$ that can contribute to a \Yf-function
$\Y{a}{s}{k}$ with $k$ negative is given by the intersection
of the diagonal $\Y{a}{s+i}{k+i}$ with the diagonal $\Y{a}{j}{-j}$.

\section{9-point multi-Regge limit}
\label{sec:9pt_bethe}
Here we present the equations governing the $\Yf$-system in the multi-Regge limit.
In the lower half of the fundamental strip $\mathrm{Im} \theta \in\left(-\pi/4,0\right)$ the equations are actually obtained simply by dropping the integral
contributions from the original expressions \eqref{eq:ysys1}-\eqref{eq:ysys3}.
For the upper half of the fundamental strip $\mathrm{Im} \theta \in\left(0,\pi/4\right)$, some residue terms survive in the multi-Regge limit.
The relevant equations take the form $\log \Yf_{a,s} = p_{a,s}$ with
the right hand side given by
\begin{flalign*}
&p_{2 \pm 1,s}\left(\theta\right)=
 -m_s\cosh\left(\theta-i\phi_s \right)\pm C_s &\\[2mm]
&p_{2,s}\left(\theta\right)\ = \ -\sqrt{2}m_s\cosh\left(\theta-i\phi_s\right)
\end{flalign*}
for the first two values $s = 1,2$ of the parameter $s$. Recall that
the corresponding phases $\phi_s$ are given by $\phi_1 = 0$ and $\phi_2
= -\pi/4$. All the remaining functions contain residue terms. These are:
\begin{flalign*}
&p_{2\pm 1,3}\left(\theta\right)=
-m_3\cosh\left(\theta+i\frac{\pi}{2}\right)\pm C_3+\log\left(1+e^{\pm C_4-m_4\cosh\left(\theta+i\frac{\pi}{2}\right)}\right)&\\[2mm]
&p_{2,3}\left(\theta\right)\ =\ -\sqrt{2}m_3\cosh\left(\theta+i\frac{\pi}{2}\right)
+\log\left(1+e^{-\sqrt{2}m_4\cosh\left(\theta+i\frac{\pi}{2}\right)}\right)&
\end{flalign*}
\begin{flalign*}
&p_{2\pm 1 ,4}\left(\theta\right)=- m_4\cosh\left(\theta+ i \frac{3 \pi}{4}\right)\pm C_4-\log\left(1+e^{-\sqrt{2}m_4\cosh\left(\theta+i\frac{\pi}{2}\right)}\right)&\\[2mm]
&p_{2,4}\left(\theta\right)=-\sqrt{2}m_4\cosh\left(\theta+i\frac{3\pi}{4}\right)-\log
\left(\left(1+e^{C_4-m_4\cosh\left(\theta+i\frac{\pi}{2}\right)}\right)
\left(1+e^{-C_4-m_4\cosh\left(\theta+i\frac{\pi}{2}\right)}\right)\right)
\end{flalign*}
As mentioned in the text, these equations could be used as the starting point to
find the solutions of equation \eqref{TBABA} numerically. If we use the above functions
$p_{a,s}$, however, the left-hand side of eqs.\ \eqref{TBABA} involve sums of products
of exponentials such as
$$ e^{p_{2,3}(\theta)} =  e^{-\sqrt{2}m_3\cosh\left(\theta+i\frac{\pi}{2}\right)}
\left(1+e^{-\sqrt{2}m_4\cosh\left(\theta+i\frac{\pi}{2}\right)}\right)\ .  $$
In order for the Bethe ansatz equations to take a more standard form one needs to work
with a larger set of momenta $p_A$ and the corresponding $S$ matrices as discussed
in the main text.
\end{appendix}


\begin{thebibliography}{9}

\bibitem{Drummond:2007au}
J.~M. Drummond, J.~Henn, G.~P. Korchemsky, and E.~Sokatchev, {\it {Conformal
  Ward identities for Wilson loops and a test of the duality with gluon
  amplitudes}},  {\em Nucl. Phys.} {\bf B826} (2010) 337--364,
  [\href{http://xxx.lanl.gov/abs/0712.1223}{{\tt arXiv:0712.1223}}].

\bibitem{Maldacena:1997re}
J.~M. Maldacena, {\it {The large N limit of superconformal field theories and
  supergravity}},  {\em Adv. Theor. Math. Phys.} {\bf 2} (1998) 231--252,
  [\href{http://xxx.lanl.gov/abs/hep-th/9711200}{{\tt hep-th/9711200}}].

\bibitem{Bern:2005iz}
Z.~Bern, L.~J. Dixon, and V.~A. Smirnov, {\it {Iteration of planar amplitudes
  in maximally supersymmetric Yang-Mills theory at three loops and beyond}},
  {\em Phys. Rev.} {\bf D72} (2005) 085001,
  [\href{http://xxx.lanl.gov/abs/hep-th/0505205}{{\tt hep-th/0505205}}].

\bibitem{Alday:2007hr}
L.~F. Alday and J.~M. Maldacena, {\it {Gluon scattering amplitudes at strong
  coupling}},  {\em JHEP} {\bf 06} (2007) 064,
  [\href{http://xxx.lanl.gov/abs/0705.0303}{{\tt arXiv:0705.0303}}].

\bibitem{Drummond:2007bm}
J.~M. Drummond, J.~Henn, G.~P. Korchemsky, and E.~Sokatchev, {\it {The hexagon
  Wilson loop and the BDS ansatz for the six- gluon amplitude}},  {\em Phys.
  Lett.} {\bf B662} (2008) 456--460,
  [\href{http://xxx.lanl.gov/abs/0712.4138}{{\tt arXiv:0712.4138}}].

\bibitem{Bern:2008ap}
Z.~Bern {\em et.~al.}, {\it {The Two-Loop Six-Gluon MHV Amplitude in Maximally
  Supersymmetric Yang-Mills Theory}},  {\em Phys. Rev.} {\bf D78} (2008)
  045007, [\href{http://xxx.lanl.gov/abs/0803.1465}{{\tt arXiv:0803.1465}}].

\bibitem{Bartels:2008ce}
J.~Bartels, L.~N. Lipatov, and A.~Sabio~Vera, {\it {BFKL Pomeron, Reggeized
  gluons and Bern-Dixon-Smirnov amplitudes}},  {\em Phys. Rev.} {\bf D80}
  (2009) 045002, [\href{http://xxx.lanl.gov/abs/0802.2065}{{\tt
  arXiv:0802.2065}}].

\bibitem{Bartels:2008sc}
J.~Bartels, L.~N. Lipatov, and A.~Sabio~Vera, {\it {\N4 supersymmetric Yang
  Mills scattering amplitudes at high energies: the Regge cut contribution}},
  {\em Eur. Phys. J.} {\bf C65} (2010) 587--605,
  [\href{http://xxx.lanl.gov/abs/0807.0894}{{\tt arXiv:0807.0894}}].

\bibitem{Goncharov:2010jf}
A.~B. Goncharov, M.~Spradlin, C.~Vergu, and A.~Volovich, {\it {Classical
  Polylogarithms for Amplitudes and Wilson Loops}},
  [\href{http://xxx.lanl.gov/abs/1006.5703}{{\tt arXiv:1006.5703}}].

\bibitem{DelDuca:2009au}
V.~Del~Duca, C.~Duhr, and V.~A. Smirnov, {\it {An Analytic Result for the
  Two-Loop Hexagon Wilson Loop in N = 4 SYM}},  {\em JHEP} {\bf 03} (2010) 099,
  [\href{http://xxx.lanl.gov/abs/0911.5332}{{\tt arXiv:0911.5332}}].

\bibitem{DelDuca:2010zg}
V.~Del~Duca, C.~Duhr, and V.~A. Smirnov, {\it {The Two-Loop Hexagon Wilson Loop
  in N = 4 SYM}},  {\em JHEP} {\bf 05} (2010) 084,
  [\href{http://xxx.lanl.gov/abs/1003.1702}{{\tt arXiv:1003.1702}}].

\bibitem{Lipatov:2010qg}
L.~N. Lipatov and A.~Prygarin, {\it {Mandelstam cuts and light-like Wilson
  loops in \N4 SUSY}},  [\href{http://xxx.lanl.gov/abs/1008.1016}{{\tt
  arXiv:1008.1016}}].

\bibitem{Dixon:2011pw}
  L.~J.~Dixon, J.~M.~Drummond and J.~M.~Henn,
  JHEP {\bf 1111} (2011) 023,
  [\href{http://xxx.lanl.gov/abs/1108.4461}{{\tt arXiv:1108.4461}}].

\bibitem{CaronHuot:2011kk}
  S.~Caron-Huot and S.~He,
  {\it Jumpstarting the All-Loop S-Matrix of Planar N=4 Super Yang-Mills},
  [\href{http://xxx.lanl.gov/abs/1112.1060}{{\tt arXiv:1112.1060}}].

\bibitem{Dixon:2012yy}
  L.~J.~Dixon, C.~Duhr and J.~Pennington,
  {\it Single-valued harmonic polylogarithms and the multi-Regge limit},
  [\href{http://xxx.lanl.gov/abs/1207.0186}{{\tt arXiv:1207.0186}}].

\bibitem{DDP}
L.~J.~Dixon, C.~Duhr and J.~Penningten, to appear.

\bibitem{Fadin:2011we}
  V.~S.~Fadin and L.~N.~Lipatov,
  Phys.\ Lett.\ B {\bf 706} (2012) 470,
  [\href{http://xxx.lanl.gov/abs/1111.0782}{{\tt arXiv:1111.0782}}].

\bibitem{Lipatov:2010ad}
  L.~N.~Lipatov and A.~Prygarin,
  {\it BFKL approach and six-particle MHV amplitude in N=4 super Yang-Mills}
  Phys.\ Rev.\ D {\bf 83} (2011) 125001,
  [\href{http://xxx.lanl.gov/abs/1011.2673}{{\tt arXiv:1011.2673}}].

\bibitem{Bartels:2011ge}
  J.~Bartels, A.~Kormilitzin, L.~N.~Lipatov and A.~Prygarin,
  [\href{http://xxx.lanl.gov/abs/1112.6366}{{\tt arXiv:1112.6366}}].

\bibitem{Lipatov:2009nt}
L.~N. Lipatov, {\it {Integrability of scattering amplitudes in \N4 SUSY}},
  {\em J. Phys.} {\bf A42} (2009) 304020,
  [\href{http://xxx.lanl.gov/abs/0902.1444}{{\tt arXiv:0902.1444}}].

\bibitem{Alday:2009yn}
L.~F. Alday and J.~Maldacena, {\it {Null polygonal Wilson loops and minimal
  surfaces in Anti- de-Sitter space}},  {\em JHEP} {\bf 11} (2009) 082,
  [\href{http://xxx.lanl.gov/abs/0904.0663}{{\tt arXiv:0904.0663}}].

\bibitem{Alday:2009dv}
L.~F. Alday, D.~Gaiotto, and J.~Maldacena, {\it {Thermodynamic Bubble Ansatz}},
   [\href{http://xxx.lanl.gov/abs/0911.4708}{{\tt arXiv:0911.4708}}].

\bibitem{Alday:2010vh}
L.~F. Alday, J.~Maldacena, A.~Sever, and P.~Vieira, {\it {Y-system for
  Scattering Amplitudes}},  \href{http://xxx.lanl.gov/abs/1002.2459}{{\tt
  arXiv:1002.2459}}.

\bibitem{Bartels:2010ej}
J.~Bartels, J.~Kotanski and V.~Schomerus, {\it{Excited Hexagon Wilson Loops
for Strongly Coupled \N4 SYM}}, {\em JHEP}{\bf 1101} (2011) 096,
[\href{http://xxx.lanl.gov/abs/1009.3938}{{\tt arXiv:1009.3938}}].

\bibitem{Dorey:1996re}
P.~Dorey and R.~Tateo, {\it {Excited states by analytic continuation of TBA
  equations}},  {\em Nucl. Phys.} {\bf B482} (1996) 639--659,
  [\href{http://xxx.lanl.gov/abs/hep-th/9607167}{{\tt hep-th/9607167}}].

\bibitem{Dorey:1997rb}
P.~Dorey and R.~Tateo, {\it {Excited states in some simple perturbed conformal
  field theories}},  {\em Nucl. Phys.} {\bf B515} (1998) 575--623,
  [\href{http://xxx.lanl.gov/abs/hep-th/9706140}{{\tt hep-th/9706140}}].

\bibitem{Hatsuda:2010vr}
Y.~Hatsuda, K.~Ito, K.~Sakai, and Y.~Satoh, {\it {Six-point gluon scattering
  amplitudes from $Z_4$-symmetric integrable model}},  {\em JHEP} {\bf 09}
  (2010) 064, [\href{http://xxx.lanl.gov/abs/1005.4487}{{\tt
  arXiv:1005.4487}}].

\bibitem{Hatsuda:2011ke}
  Y.~Hatsuda, K.~Ito, K.~Sakai and Y.~Satoh,
  {\it g-functions and gluon scattering amplitudes at strong coupling},
  JHEP {\bf 1104} (2011) 100
  [\href{http://xxx.lanl.gov/abs/1102.2477}{{\tt arXiv:1102.2477}}].

\bibitem{Hatsuda:2011jn}
  Y.~Hatsuda, K.~Ito and Y.~Satoh,
  {\it T-functions and multi-gluon scattering amplitudes},
  JHEP {\bf 1202} (2012) 003
  [\href{http://xxx.lanl.gov/abs/1109.5564}{{\tt arXiv:1109.5564}}].

\bibitem{Alday:2010ku}
  L.~F. Alday, D.~Gaiotto, J.~Maldacena, A.~Sever, and P.~Vieira, {\it {An Operator Product Expansion for Polygonal null Wilson Loops}},
  \href{http://xxx.lanl.gov/abs/1006.2788}{{\tt arXiv:1006.2788}}.

\bibitem{Gaiotto:2010fk}
  D.~Gaiotto, J.~Maldacena, A.~Sever and P.~Vieira,
  {\it Bootstrapping Null Polygon Wilson Loops},
  JHEP {\bf 1103} (2011) 092
  [\href{http://xxx.lanl.gov/abs/1010.5009}{{\tt arXiv:1010.5009}}].

\bibitem{Gaiotto:2011dt}
  D.~Gaiotto, J.~Maldacena, A.~Sever and P.~Vieira,
  JHEP {\bf 1112} (2011) 011
  [\href{http://xxx.lanl.gov/abs/1102.0062}{{\tt arXiv:1102.0062}}].

\bibitem{Bartels:2011xy}
  J.~Bartels, L.~N.~Lipatov and A.~Prygarin,
  {\it Collinear and Regge behavior of $2\rightarrow 4$ MHV amplitude in N = 4 super Yang-Mills theory},
  arXiv:1104.4709 [hep-th].

\bibitem{Brower:1974yv}
  R.~C.~Brower, C.~E.~DeTar and J.~H.~Weis,
  {\it Regge Theory for Multiparticle Amplitudes}
  Phys.\ Rept.\  {\bf 14} (1974) 257.

\bibitem{Yang2011}
G.~Yang, {\it {A simple collinear limit of scattering amplitudes
at strong coupling}}, {\em JHEP}, {\bf 1103} (2011) 87
[\href{http://xxx.lanl.gov/abs/1006.3306}{{\tt arXiv:1006.3306}}].

\bibitem{Gaiotto:2009hg}
  D.~Gaiotto, G.~W.~Moore and A.~Neitzke,
  {\it Wall-crossing, Hitchin Systems, and the WKB Approximation},
  [\href{http://xxx.lanl.gov/abs/0907.3987}{{\tt arXiv:0907.3987}}].

\bibitem{Gaiotto:2012rg}
  D.~Gaiotto, G.~W.~Moore and A.~Neitzke,
  {\it Spectral networks},
  [\href{http://xxx.lanl.gov/abs/1204.4824}{{\tt arXiv:1204.4824}}].

\bibitem{BKP}
J.~Bartels, Nucl.\ Phys.\ B {\bf 175} (1980) 365,\\
J.~Kwiecinski and M.~Praszalowicz, Phys.\ Lett.\ B {\bf 94} (1980) 413.

\bibitem{Bartels:2011nz}
  J.~Bartels, L.~N.~Lipatov and A.~Prygarin,
  {\it Integrable spin chains and scattering amplitudes},
 J.\ Phys.\ A A {\bf 44} (2011) 454013,
 [\href{http://xxx.lanl.gov/abs/1104.0816}{{\tt arXiv:1104.0816}}].

\end{thebibliography}
\end{document}